\documentclass{article}

\PassOptionsToPackage{numbers, compress}{natbib}

\usepackage[final]{neurips_2023}

\usepackage[utf8]{inputenc} 
\usepackage[T1]{fontenc}    
\usepackage{hyperref}       
\usepackage{url}            
\usepackage{booktabs}       
\usepackage{amsfonts}       
\usepackage{nicefrac}       
\usepackage{microtype}      
\usepackage{xcolor}         
\usepackage{enumitem}
\usepackage{graphicx}
\usepackage{float}
\usepackage{wrapfig}
\usepackage[most]{tcolorbox}
\usepackage{xcolor}
\usepackage{amssymb}
\usepackage{hyperref}
\let\svthefootnote\thefootnote
\newcommand\freefootnote[1]{%
  \let\thefootnote\relax%
  \footnotetext{#1}%
  \let\thefootnote\svthefootnote%
}

\title{Democratic Policy Development

using Collective Dialogues and AI}

\author{%
  Andrew Konya\thanks{Corresponding author: andrew@remesh.org}\\
  Remesh\\
\And
  Lisa Schirch \\
  University of\\ Notre Dame \\
\And
   Colin Irwin \\
   University of \\ Liverpool \\
\And
   Aviv Ovadya\\
    AI \& Democracy \\ Foundation \\
}

\begin{document}

\maketitle
\vspace{1cm}
\begin{abstract}
   We design and test an efficient democratic process for developing policies that reflect informed public will. The process combines AI-enabled collective dialogues that make deliberation democratically viable at scale with bridging-based ranking for automated consensus discovery. A GPT4-powered pipeline translates points of consensus into representative policy clauses from which an initial policy is assembled. The initial policy is iteratively refined with the input of experts and the public before a final vote and evaluation. We test the process three times with the US public, developing policy guidelines for AI assistants related to medical advice, vaccine information, and wars \& conflicts. We show the process can be run in two weeks with 1500+ participants for around \$10,000, and that it generates policy guidelines with strong public support across demographic divides. We measure 75-81\% support for the policy guidelines overall, and no less than 70-75\% support across demographic splits spanning age, gender, religion, race, education, and political party. Overall, this work demonstrates an end-to-end proof of concept for a process we believe can help AI labs develop common-ground policies, governing bodies break political gridlock, and diplomats accelerate peace deals.
\end{abstract}

\freefootnote{Author contributions: Andrew developed the AI tools used in the process. Andrew, Colin, and Lisa designed and tested the process. Aviv advised on process design and implementation. Everyone contributed to this report.}

\freefootnote{Code and data available at: \url{https://github.com/openai/democratic-inputs/tree/main/projects/collective_dialogues_for_democratic_input}}
\newpage

\section*{Executive Summary}

\textbf{We introduce a democratic process for developing policies that reflect informed public will}. The process integrates democratic inputs with subject matter expertise to yield policies optimized for both representativeness and quality. AI-augmented collective dialogues make deliberation democratically viable at scale. Bridging-based ranking enables rapid consensus discovery. GPT4-powered tools make the process efficient. Modularization makes the process reproducible.  

\textbf{At the heart of this process are \emph{collective dialogues}} on Remesh which involve participants first being educated on an issue, followed by structured text-based deliberation where participants iteratively: a) respond to open-ended prompts, b) see and evaluate each other’s responses, and c) reflect on representative perspectives. To make democratic representation possible at scale, every participant's agreement with every response is approximated from sparse evaluations using elicitation inference.

\textbf{Process overview}:

\begin{enumerate}
    \item \textbf{Learn public views}: An initial collective dialogue elicits informed perspectives from a carefully selected representative public.
    \item \textbf{Create initial policy}: Bridging-based ranking is used to identify points of consensus elicited during the collective dialogue. A GPT4-powered pipeline rapidly translates points of consensus into representative policy clauses from which an initial policy is assembled. 
    \item \textbf{Expert refinement}: Relevant experts refine the policy into a higher-quality version that incorporates specialists' knowledge, minimizes ambiguities, and better handles edge cases.
    \item \textbf{Public refinement}: The policy is further refined to be more representative through a second collective dialogue with a representative public.
    \item \textbf{Evaluation}: Public support for the final policy is assessed via a third collective dialogue with a large-scale, highly representative public. Consistency with precedent policy is estimated using GPT4.
\end{enumerate}
 
 \textbf{We tested the process three times}, developing policy guidelines for AI assistants related to situations involving \emph{medical advice}, \emph{vaccine information}, and \emph{wars \& conflicts}. \emph{Each process run} took two weeks, cost on the order of \$10,000 USD, and included democratic input from 1500+ participants representative of the US population; including around 5000 text responses and nearly 100,000 votes. The resulting policy guidelines had strong support across US demographic divides: between 75-81\% public support overall, and no less than 70-75\% support across demographic groups spanning age, gender, religion, race, education, and political party. Zero \emph{conflicts} between the policy guidelines and the Universal Declaration of Human Rights were found, and a \emph{consistent} relationship was estimated between individual rights and guideline clauses between 13-27\% of the time (the rest being
\emph{neutral}). 

\textbf{AI labs and governing bodies can use this process} to develop concise sets of common ground policy guidelines that bridge demographic divides and reflect what a given population wants. It is ideal for those that have to make policy decisions that impact large populations and want a democratic process to align those decisions with informed public will. However, while we view the work presented here as an end-to-end proof of concept that can be used today, it is not a perfect polished system. Every aspect of every step of the process can be critiqued and improved. 

\textbf{Future work will focus on refining the process by using it.} We aim to use the process to help AI labs develop common-ground policies, governing bodies break political gridlock, and diplomats accelerate peace deals. To refine the process we aim to iterate based on the needs arising from real-world use. We expect this will lead to things like developing better process tooling, increasing process standardization, accommodating more complex policy formats, integrating objective policy quality metrics, and developing approaches to efficiently recruit globally representative participants.

\newpage

\section{Motivation}
\textbf{We aim to create policies that reflect informed public will}. Our core motivation is to increase the probability that the future aligns with the will of humanity. We see aligning the behavior of high-impact systems---from governments to AGI--- with informed public will as instrumental to this goal. \emph{Policies} specify desired system behaviors in ways that can be practically implemented. We thus focus on creating an approach for developing policies that reflect informed public will. In this pursuit, two critical challenges arise. First, public will constitutes a plurality of views about how a system should behave, with some views in direct conflict. Which views should ultimately be reflected in policies? Second, developing quality policies on most issues requires some expertise the general public lacks. How do you create policy that reflects public will yet integrates expertise?

\textbf{We draw inspiration from peace negotiations and citizens' assemblies} to address the first challenge---identifying which public views should be reflected in a policy. During peace negotiations, the challenge is to find points of common ground between conflicting parties that align with everyone's interests \cite{manual2010peck,guidence2012united}. While final agreements often go beyond points of consensus to include trade-offs between sides, identifying common ground often forms the basis for making more complicated issues surmountable. However, finding common ground typically requires a shared understanding of reality which the public often lacks, especially on contentious issues. Citizens’ assemblies \cite{oecd2020innovative} create this shared understanding by providing participants with a balanced education on an issue and fostering deliberation among them to help understand other’s views. But, citizen assemblies can take months to coordinate and cost hundreds of thousands or millions of dollars to execute. Technology-enabled collective dialogues ---which are increasingly common in peacebuilding \cite{UN2021williams,irwin2021using,UN2021jeanine,UN2020cutting,UN2022lynn,UN2023carol,UN2023liita,bilich2019faster}---offer similar affordances, yet can be executed in hours or days for thousands of dollars by just a few people. Thus, we design our process to generate policies that reflect informed public consensus and leverage collective dialogues to educate participants, foster deliberation, and elicit informed views from which consensus can be identified. 

\textbf{We conceptualize policy development as an optimization process along two axes} to address the second challenge---integrating expertise and public will. The first axis is \emph{representativeness} and captures how well the policy reflects informed public consensus on an issue. The second axis is \emph{quality} and captures the degree to which the policy is clear, unambiguous, and reflects expert knowledge. While these axes oversimplify the complex universe of desiderata one might ascribe to policy, they help factor the process in a practical way. Increasing representativeness comes from public input. Increasing quality comes from expert input. We thus design a process that iterates between public and expert input to produce policies that are high in both quality and representativeness.

\section{Collective Dialogues}

\begin{figure}[H]
\vspace{-1.5cm}
\hspace{-1cm}
  \includegraphics[width=1.1\linewidth]{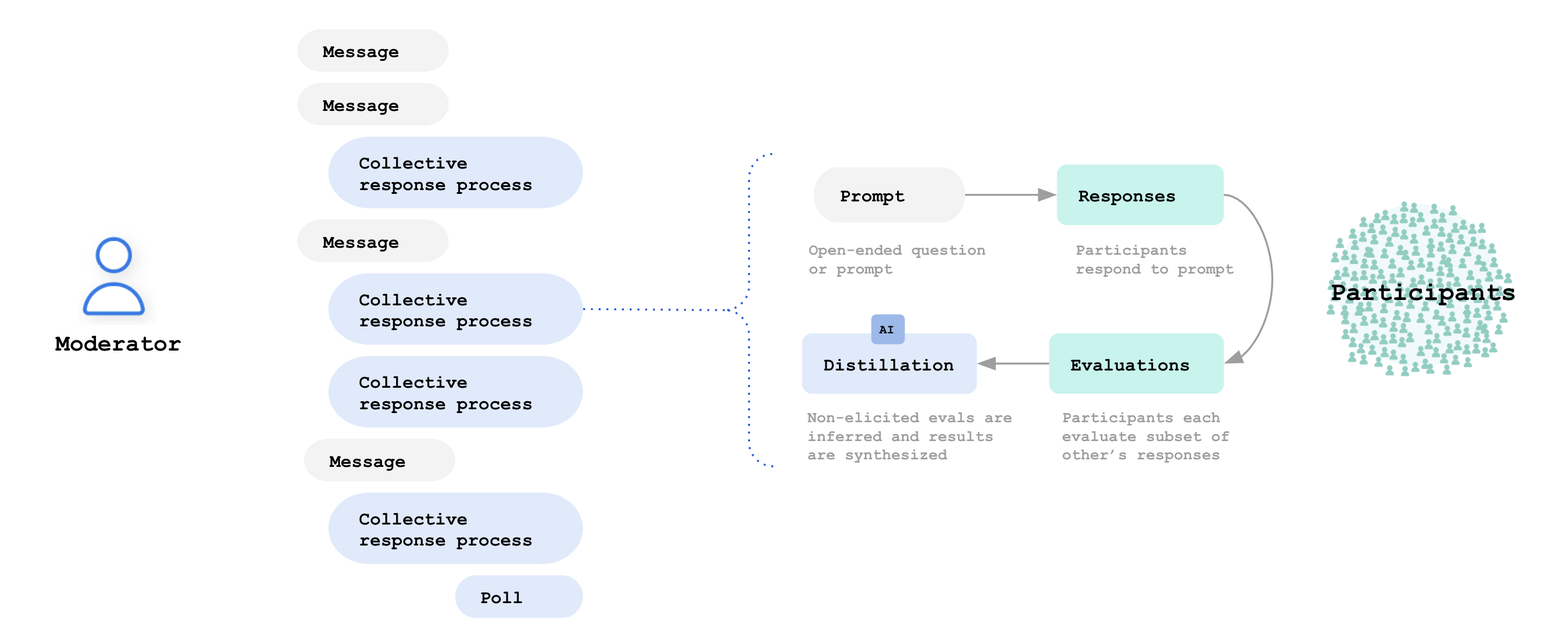}
  \caption{Diagram showing the key elements of a collective dialogue. A moderator guides a back-and-forth exchange with a group which involves sending messages, polls, and triggering collective response processes. During each collective response process, participants respond to an open-ended prompt then see and evaluate the responses of others. Then representative results are distilled and shared with the moderator and (potentially) the participants.}
  \label{fig:CD diagram}
\end{figure}

\begin{figure}[H]
\centering
  \includegraphics[width=1.0\linewidth]{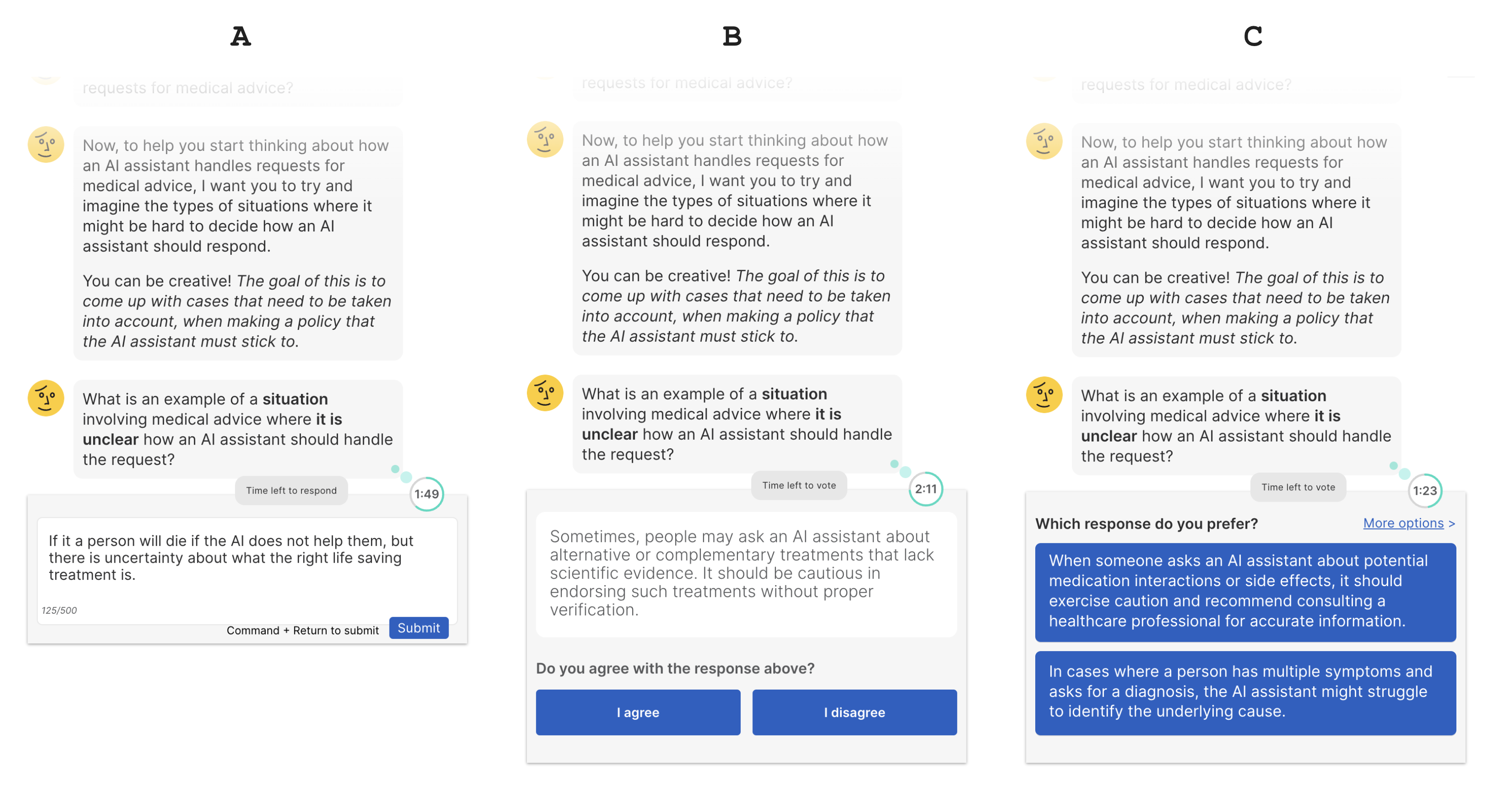}
  \caption{Screenshots showing the key participatory actions for each collective response process that takes place during a collective dialogue on Remesh: A) Submit natural language response to prompt, B) review others' responses and vote if you agree, C) review pairs of other's responses and vote which you prefer.}
  \label{fig:CRS participation}
\end{figure}

\textbf{A collective dialogue process is an iterative back-and-forth exchange} between a moderator and participants. During each turn of the dialogue, participants are sent either a read-only message (text, image, or video), a poll, or an open-ended prompt that kicks off a \emph{collective response process} (figure \ref{fig:CD diagram}). During a collective response process\cite{ovadya2023generative}, participants first share a natural language response to the prompt, then they evaluate the responses submitted by others. The evaluation step serves two purposes: first, it exposes participants to other’s views to help them understand each other, and second, it elicits the data needed to quantify response representativeness and identify points of consensus. On the collective dialogue platform Remesh\footnote{\url{https://www.remesh.ai/product}}, two types of evaluations are elicited; agreement votes, and pair choice votes (figure \ref{fig:CRS participation}). However, when there are hundreds or thousands of participants, each participant is only able to vote on a small fraction of the submitted responses. Thus, elicitation inference \cite{konya2022elicitation,bilich2019faster} is used to convert a sparse vote sampling into a complete vote matrix from which aggregate results---like the overall fraction of participants who agree with a response---can be computed. 

\begin{figure}[H]
\centering
  \includegraphics[width=1.0\linewidth]{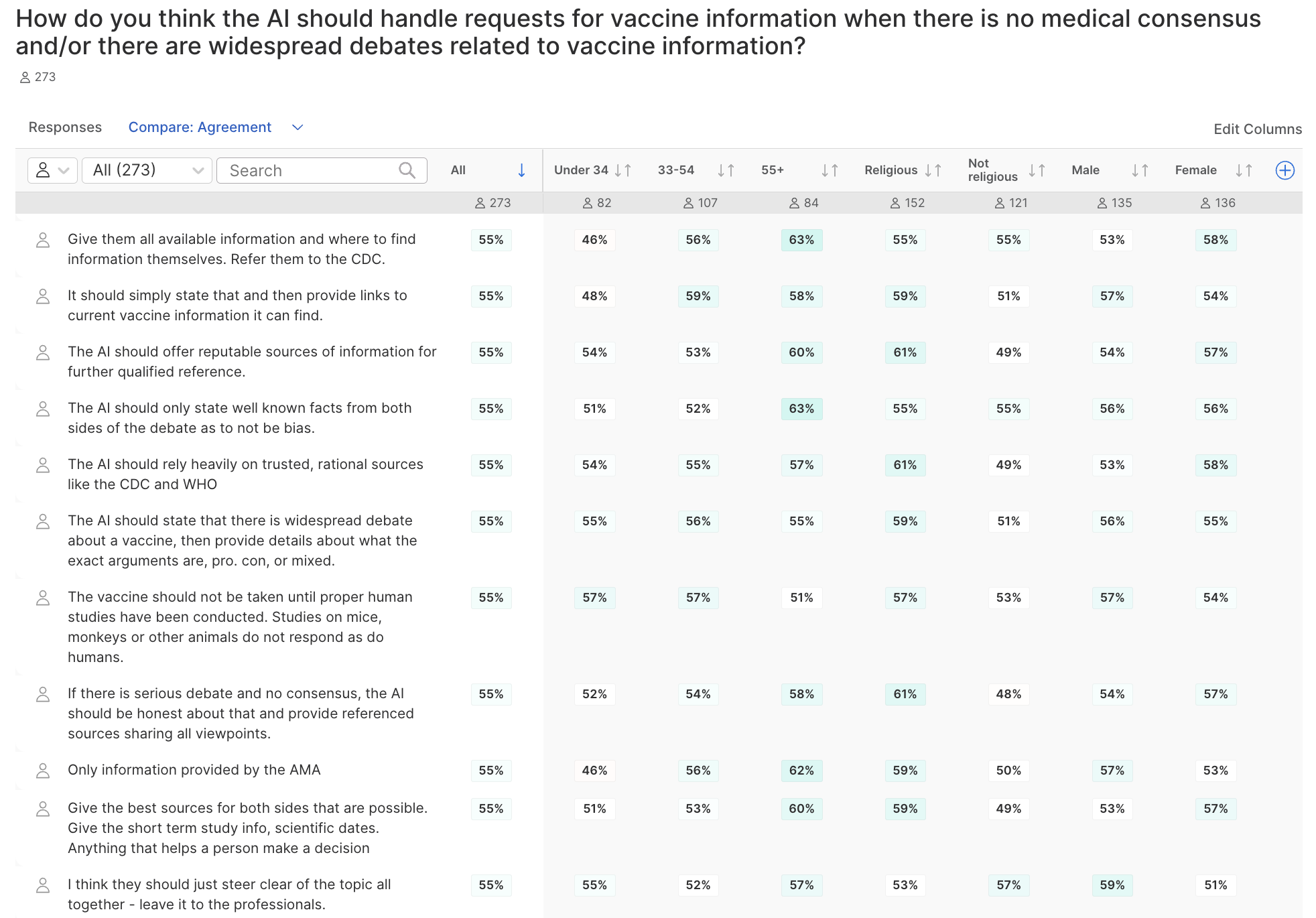}
  \caption{Remesh screenshot showing an example of the results generated from each collective response process that takes place during a collective dialogue---here the percentage of every demographic segment that agrees with each response is shown. This data is used within our process to compute \emph{bridging agreement} across demographic segments and identify points of consensus. }
  \label{fig:CRS results}
\end{figure}

\textbf{During a live collective dialogue process, all participation is simultaneous} and each collective response process takes a few minutes to complete. When each collective response process is completed, every participant sees the fraction of participants agreeing with their response as well as a representative subset of responses from the group (figure \ref{fig:participant results}). The moderator sees preliminary results as each process unfolds and final results when each completes. Those results include common topics and their frequency, the fraction of participants---overall and within each demographic segment (figure \ref{fig:CRS results})---who agree with each response, and a plural subset of responses selected to include at least one response that each participant prefers over most others. Based on what the moderator learns from these results, they can either continue the dialogue based on their pre-programmed discussion guide\footnote{The \emph{discussion guide} captures the planned sequence of messages, polls, and collective response prompts for a collective dialogue.} or pivot in real-time to drill deeper into an issue.

\textbf{A collective dialogue process can also be run asynchronously}. In this case, the dialogue simply follows the pre-programmed discussion guide, with each participant starting and completing the dialogue on their own timeline. During each collective response process within the dialogue, participants still evaluate the responses of the participants who came before them, but they do not see the overall agreement with their response nor the set of representative responses. When the dialogue concludes, the same set of results are available regardless of the modality being asynchronous or live. Overall, an asynchronous collective dialogue is logistically easier to execute, but it comes at the cost of the flexibility of live moderation and the convergent feedback loop of participants seeing live representative results. We experiment with using both modalities in our policy development process. 

\section{Democratic process}

\begin{figure}[H]
\centering
\hspace*{-2cm} 
  \includegraphics[width=1.3\linewidth]{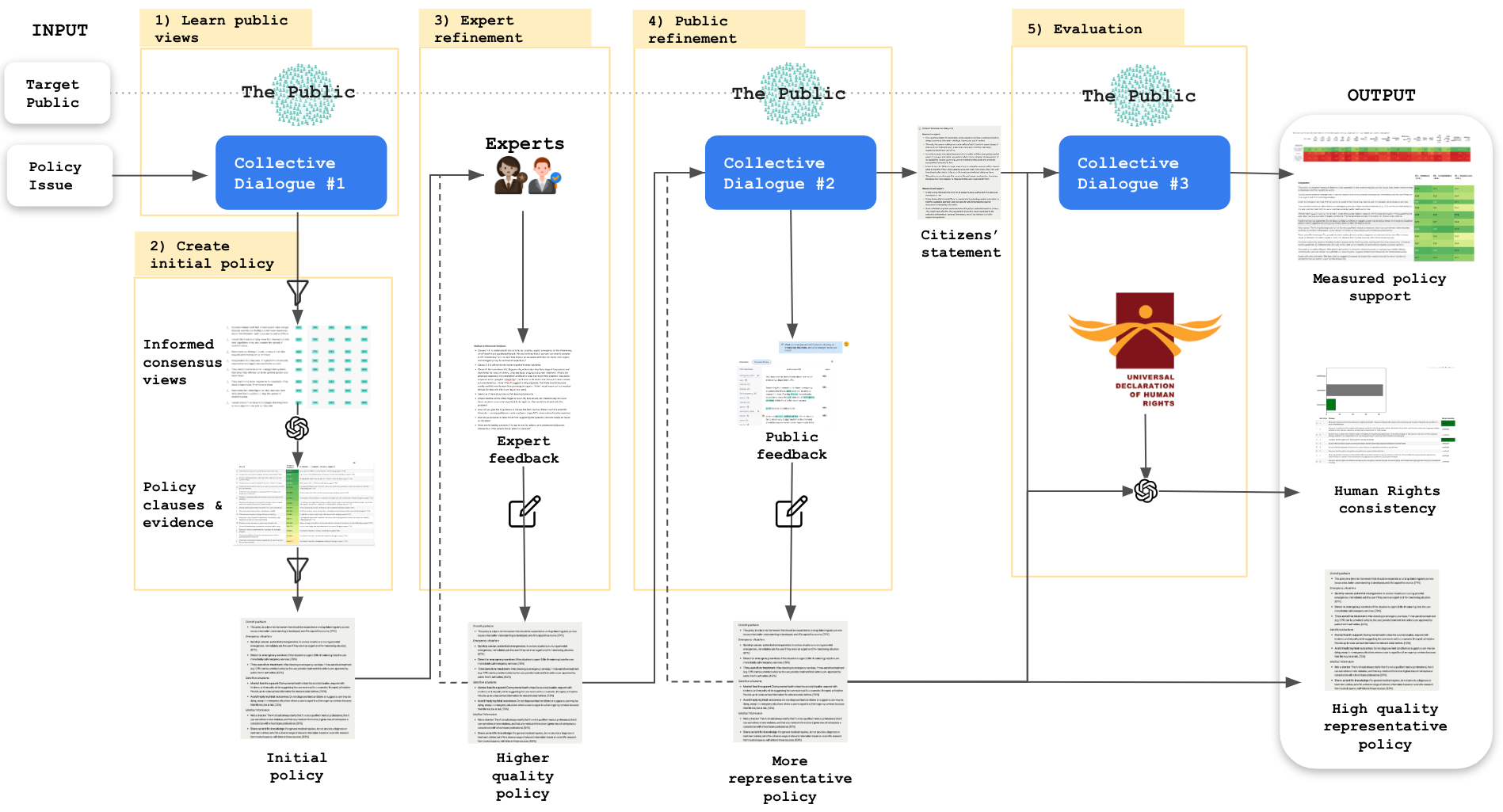}
  \caption{Diagram showing the full deliberative process. Steps 1\&2  generate an initial policy that reflects informed public consensus. In step 3 policy quality is increased with expert input. In step 4 policy representativeness is increased with public input. In step 5 the final policy is evaluated.}
  \label{fig:process}
\end{figure}

\textbf{The democratic process} takes a \emph{policy issue} and \emph{target public} as input and outputs a quality representative policy along with the public’s support for the policy and its consistency with precedent policy like the Universal Declaration of Human Rights (figure \ref{fig:process}). The process combines collective dialogues (via Remesh) to scale public deliberation with bridging-based ranking to identify points of consensus. A GPT4-powered pipeline rapidly translates points of consensus into representative policy clauses from which an initial policy is assembled. Experts refine the policy to produce a higher-quality version. The policy is further refined to be more representative through another collective dialogue with the public. Then the final policy is evaluated. Support for the final policy is assessed through a collective dialogue with a legitimately representative public, and consistency with precedent policy is estimated using GPT4.   Before the process begins, two decisions must be made:
\begin{itemize}
    \item \textbf{Policy issue}: What issue will the process develop policy to address? This can be specified in the form of a question, ie. “How should AI assistants handle requests for medical advice?” 
    \item \textbf{Target public}: What \emph{public} will the process aim to represent in the policy it generates? This determines who the participants recruited for the process need to be representative of, ie. “US citizens,” “Humanity,” etc. 
\end{itemize}

\subsection{Learn Public Views}

\begin{tcolorbox}[colback=blue!5!white,colframe=blue!30!white]
  \emph{\textbf{Inputs}: Policy issue, target public.}

\emph{\textbf{Outputs}: Collective dialogue data containing informed public views on policy issue.}
\end{tcolorbox}

A collective dialogue is run via Remesh with a group of participants selected to be representative of the target public. During the collective dialogue, participants learn about the policy issue, deliberate the issue, and then their (now informed) views on the issue are elicited. The facilitation team executes the following:

\begin{enumerate}
    \item \textbf{Create a discussion guide} on Remesh for the collective dialogue to follow which gathers appropriate demographics, sets context and educates participants on the issue, facilitates deliberation around the issue, and finally elicits participants' views on the issue (\ref{A:learn public views}.
    \item \textbf{Recruit a representative public} to participate in the collective dialogue by selecting a set of participants whose demographic distribution matches that of the target population. For example, we recruited sets of 200-300 participants representing the US public using census-balanced sampling techniques implemented on Prolific.
    \item \textbf{Moderate the collective dialogue\footnote{This type of real-time moderation is only required for live collective dialogues. In asynchronous collective dialogues, a moderator can still adjust the discussion while data is being collected, but only those who participate after the change will experience it.}} through a combination of sending pre-programmed items from the discussion guide and pivoting or probing to ask new questions on the fly as unexpected issues, contentions, and ideas are surfaced. 
\end{enumerate}

\subsection{Create Initial Policy}

\begin{tcolorbox}[colback=blue!5!white,colframe=blue!30!white]
\emph{\textbf{Inputs}: Collective dialogue data containing informed public views on a policy issue.}

\emph{\textbf{Outputs}: Initial policy reflecting informed public consensus.}
\end{tcolorbox}

From the views elicited during the collective dialogue, bridging-based ranking \cite{ovadya2022bridging,recommendation2022bridging} is used to automatically identify points of consensus. Then a GPT4-powered pipeline is used to rapidly translate consensus points into representative policy clauses from which an initial policy is assembled. This is executed through the following process:

\begin{enumerate}
\item For each collective response prompt in the \emph{elicitation section}:
\begin{enumerate}
    \item Select responses with the highest \emph{bridging agreement\footnote{Here we use \emph{max-min bridging agreement} as defined in \ref{A:create initial policy}. However, there are a range of different metrics that could be used to select bridging responses; for example, the \emph{group informed consensus} metric implemented on Polis \cite{small2021polis}.}}.
    \item Summarize the ideas in those responses and generate policy clauses using GPT4.
    \item For each policy clause generated:
    \begin{enumerate}
        \item Find the response most related to that clause.
        \item Estimate how well that response justifies that clause.
    \end{enumerate}
\end{enumerate}
\item Merge all policy clauses into one list and rank by strength of justification.
\item Choose a subset of generated clauses to become the initial policy
\end{enumerate}

Steps 1 \& 2 are done using an automated pipeline. Step 3 is done manually by the process facilitators.

\subsection{Expert Refinement}
\begin{tcolorbox}[colback=blue!5!white,colframe=blue!30!white]
\emph{\textbf{Inputs}: Initial policy reflecting informed public consensus.}

\emph{\textbf{Outputs}: Higher-quality policy integrating domain expertise.}
\end{tcolorbox}

The facilitation team identifies experts relevant to the policy issue and shares the initial policy with them. Those experts provide feedback on the initial policy; they offer suggested revisions that better reflect domain expertise or improve clarity, and point out gaps or edge cases the policy does not address. The facilitation team then refines the policy based on this input. This cycle of expert feedback and revision can happen multiple times, and can sometimes include automated tools for edge case analysis.

\subsection{Public Refinement}

\begin{tcolorbox}[colback=blue!5!white,colframe=blue!30!white]
\emph{\textbf{Inputs}: Higher-quality policy integrating domain expertise, target public.}

\emph{\textbf{Outputs}: High-quality representative policy, citizens’ statement}
\end{tcolorbox}

A collective dialogue is run via Remesh with a group of participants selected to be representative of the target public. During the collective dialogue, participants learn about the policy issue, review the latest version of the policy, provide feedback on it, and contribute reasons for and against supporting it. The policy is refined based on the provided feedback to produce a version that is more representative, and a sort of \emph{citizens' statement}\footnote{A 'citizens (review) statement' is typically a set of arguments for or against a policy proposal or set of recommendations generated by a representative panel of citizens via deliberation \cite{citizens2018}.} is assembled from the reasons for and against supporting it. The facilitation team executes the following.
\begin{enumerate}
    \item \textbf{Create a discussion guide} on Remesh for the collective dialogue to follow which collects participants' demographics, sets context and educates them on the issue, presents the policy, elicits feedback on the policy, and elicits arguments for or against the policy (\ref{A: public refinement}).
    \item \textbf{Recruit a representative public} to participate in the collective dialogue by selecting a set of participants whose demographic distribution matches that of the target population\footnote{This step is typically done using an asynchronous collective dialogue where live moderation is not needed.}. 
    \item \textbf{Refine the policy} based on public feedback collected during the collective dialogue. For example, one can identify the concerns of participants who do not support the policy and tweak the policy to better address them.
    \item \textbf{Assemble a `citizens statement'\footnote{The type of 'citizens statement' assembled here may not manifest the same standards of deliberative rigor as what is produced by a citizens' assembly \cite{citizens2018}.}} by selecting a diverse collection of the most agreed-upon reasons for and against supporting the policy elicited during the collective dialogue. 
\end{enumerate}

\subsection{Evaluation}
\begin{tcolorbox}[colback=blue!5!white,colframe=blue!30!white]
\emph{\textbf{Inputs}: High-quality representative policy, citizens’ statement, target public.}

\emph{\textbf{Outputs}: Legitimate measures of support for final policy, consistency with human rights.}
\end{tcolorbox}

A collective dialogue is conducted with a large-scale, highly representative public to assess measures of support for the policy.  During the collective dialogue, participants learn about the policy issue and the policy development process, review the final policy, and vote on their support for each clause and the policy overall. Additionally, consistency between the final policy and precedent policy (ie. the Universal Declaration of Human Rights) is estimated using GPT4. The facilitation team executes the following:

\begin{enumerate}
    \item \textbf{Create a discussion guide} on Remesh for the collective dialogue to follow which collects participants' demographics, sets context and educates them on the issue, presents the policy along with the "citizens' statement", and then measures their support for the policy overall as well as each of its clauses (\ref{A:evaluation}). 
    \item \textbf{Recruit a highly representative public} to participate in the collective dialogue by selecting a set of participants whose demographic distribution matches that of the target population. For example, we recruited sets of 1000 participants representing the US public for this step using census-balanced sampling techniques implemented on Prolific.
    \item \textbf{Compute support measures} from collective dialogue data, including overall support for the policy as well as bridging support for the policy (ie. the lowest support observed across a range of demographic groups). 
    \item \textbf{Estimate policy consistency} with the precedent policy like the Universal Declaration of Human Rights. To do this each clause of the policy is compared with each precedent clause and assessed via GPT4 to be either consistent, neutral, or conflicting. 
\end{enumerate}

\section{Experiments}
We ran the process outlined above three times to develop policy guidelines around how AI assistants should handle situations related to:
\begin{enumerate}
    \item \textbf{Medical advice}
    \item \textbf{Wars and conflicts}
    \item \textbf{Vaccines}
\end{enumerate}

\begin{figure}[H]
\centering
  \includegraphics[width=0.69\linewidth]{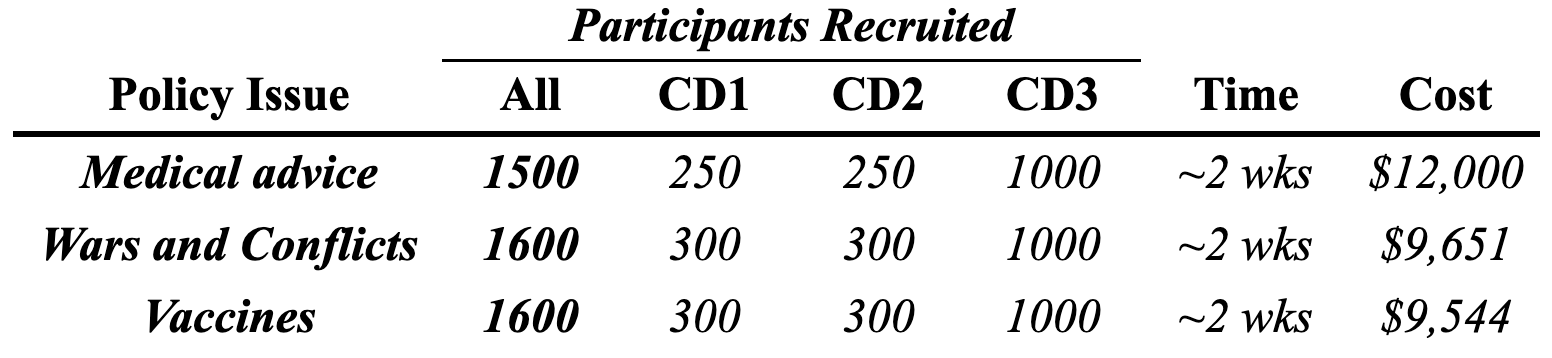}
  \caption{Summary of experiments run including the number of participants recruited for each collective dialogue, the time it took to execute the process end-to-end, and the total cost of recruiting participants for each process run in USD. Note that around 90\% of the participants recruited for each collective dialogue fully completed it.}
  \label{fig:experiments table}
\end{figure}

We chose the United States as the target public. \emph{Each process run} took around two weeks and incorporated democratic inputs from 1500+ people through collective dialogues run on Remesh (figure \ref{fig:experiments table}), including around 5000 responses to collective response prompts and nearly 100,000 votes. Participants were recruited using census-balanced sampling techniques via Prolific to be representative of the US public. Participants were paid between \$12-15 / hour USD for their time and the total cost per run was on the order of \$10,000 USD. Experts for each process run were recruited via the process facilitators' personal networks and included AI policy experts, doctors \& medical researchers, UN personnel, and participants from prior bi-partisan vaccine dialogues. 

\section{Results}

\subsection{Policy evaluation}

\begin{figure}[H]
\centering
  \includegraphics[width=0.9\linewidth]{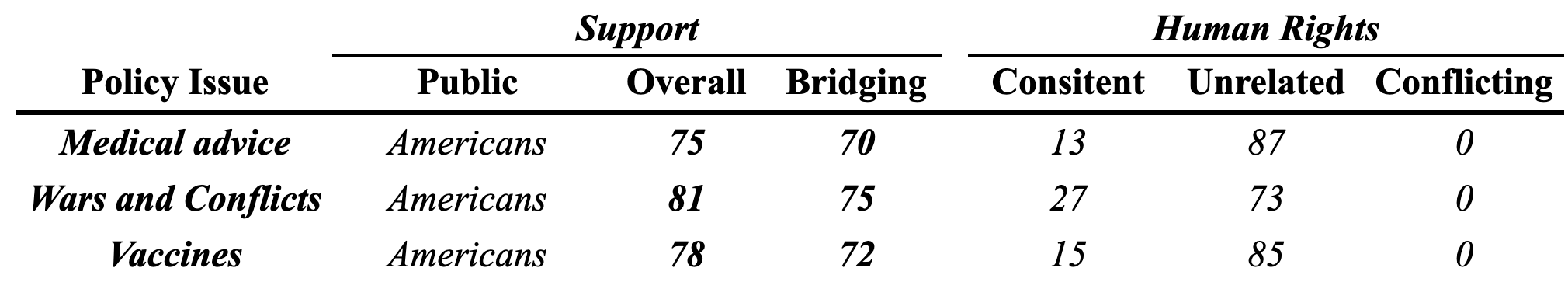}
  \caption{Evaluation results for the three policies developed. The percent \emph{overall support} measured for the policy is shown along with the \emph{bridging support} across demographic groups spanning age, gender, religion, race, education, and political party. Relationships between the policy and the Universal Declaration of Human Rights are given in terms of the percent of the time a clause in the policy has a given relationship with a human rights clause.}
  \label{fig:Policy results}
\end{figure}

The final policies generated by the process for each issue can be found in \ref{A:policies}. We measured overall support for the three policies to be between 75-81\%, and bridging support\footnote{The minimum support found within any one of a set of demographic groups.} to be between 70-75\% across demographic groups spanning age, gender, religion, race, education, and political party (figure \ref{fig:Policy results}, \ref{A:bridging support}). Overall support for individual policy clauses ranged from 63-95\%. Zero \emph{conflicts} with the Universal Declaration of Human Rights were found and a \emph{consistent} relationship was estimated between individual human rights and policy clauses between 13-27\% of the time (the rest being \emph{neutral}). Overall, we observed that the policies resulting from the process took a form closer to human-readable guidelines than intricate technical policies---here is an example clause from each policy:
\begin{itemize}
    \item \emph{\textbf{Avoid implying fatal outcomes}: Do not diagnose fatal conditions or suggest a user may be dying, except in emergency situations where a user is urged to call emergency services because their life may be at risk.}
    \item \emph{\textbf{Do not produce misinformation}: Do not generate any text, images, videos, or data sets related to conflicts which mimic the appearance of credible news, evidence, analysis, or statements by world leaders.}
    \item \emph{\textbf{Prioritize science over corporate vaccine information}: In cases of contradiction between pharmaceutical company information and medical journals, prioritize medical journals.}
\end{itemize}

\begin{figure}[H]
\hspace{-0.5cm}
  \includegraphics[width=1.1\linewidth]{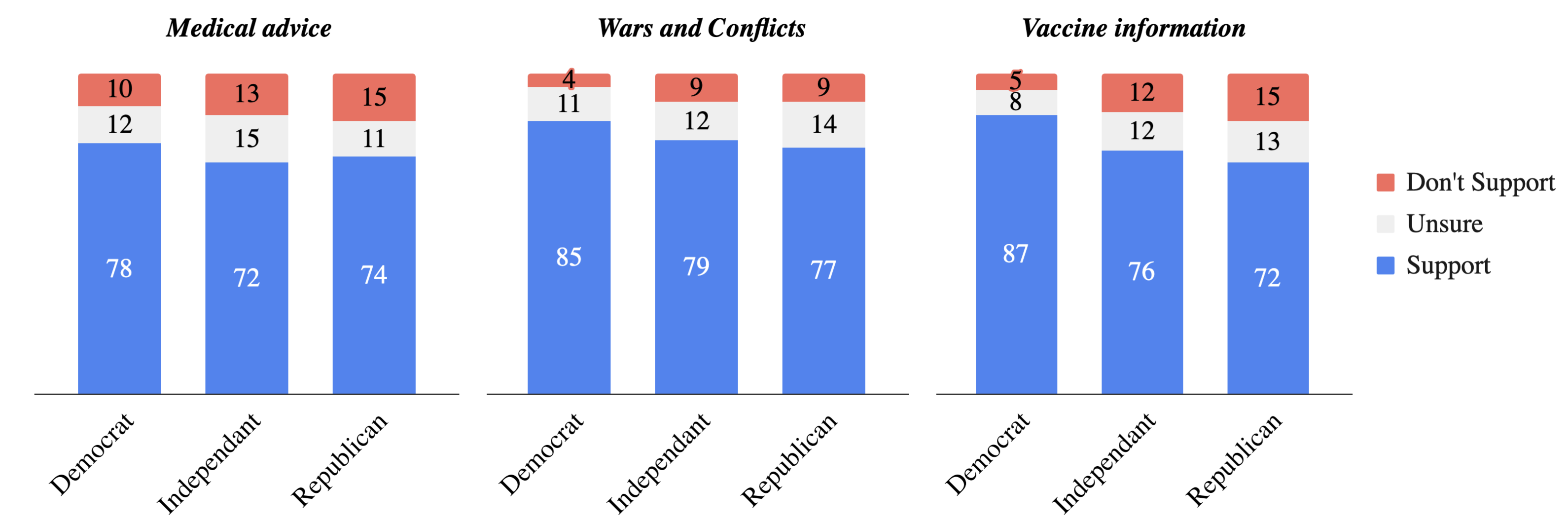}
  \caption{Support across the US political spectrum for the three different policy guidelines developed using the process.}
  \label{fig:political bridging}
\end{figure}

The motivation for running the process on \emph{vaccine information} was to pressure test how well the process could find consensus and generate bridging policy guidelines around a topic with significant political division. All policy guidelines our process generated---including on vaccine information---had 72\%+ support across Democrats, Independents, and Republicans (figure \ref{fig:political bridging}). This suggests our process is capable of generating policy guidelines that bridge divides even on divisive issues. 

\subsection{Evidence for deliberative state change}
\begin{figure}[H]
\centering
\begin{minipage}{.35\textwidth}
  \centering
  \caption{Evidence of deliberative state change. A) Shows the percent of participants who said \emph{Yes, I think the public has insight useful to guide how AI Assistants answer difficult questions} before and after participating in collective dialogues 1 or 2.  B) Shows the percentage of participants who said they supported a policy before and after evaluating each individual clause in the policy.}
  \label{fig:state change}
\end{minipage}
\hfill
\begin{minipage}{.63\textwidth}
  \centering
   \includegraphics[width=1\linewidth]{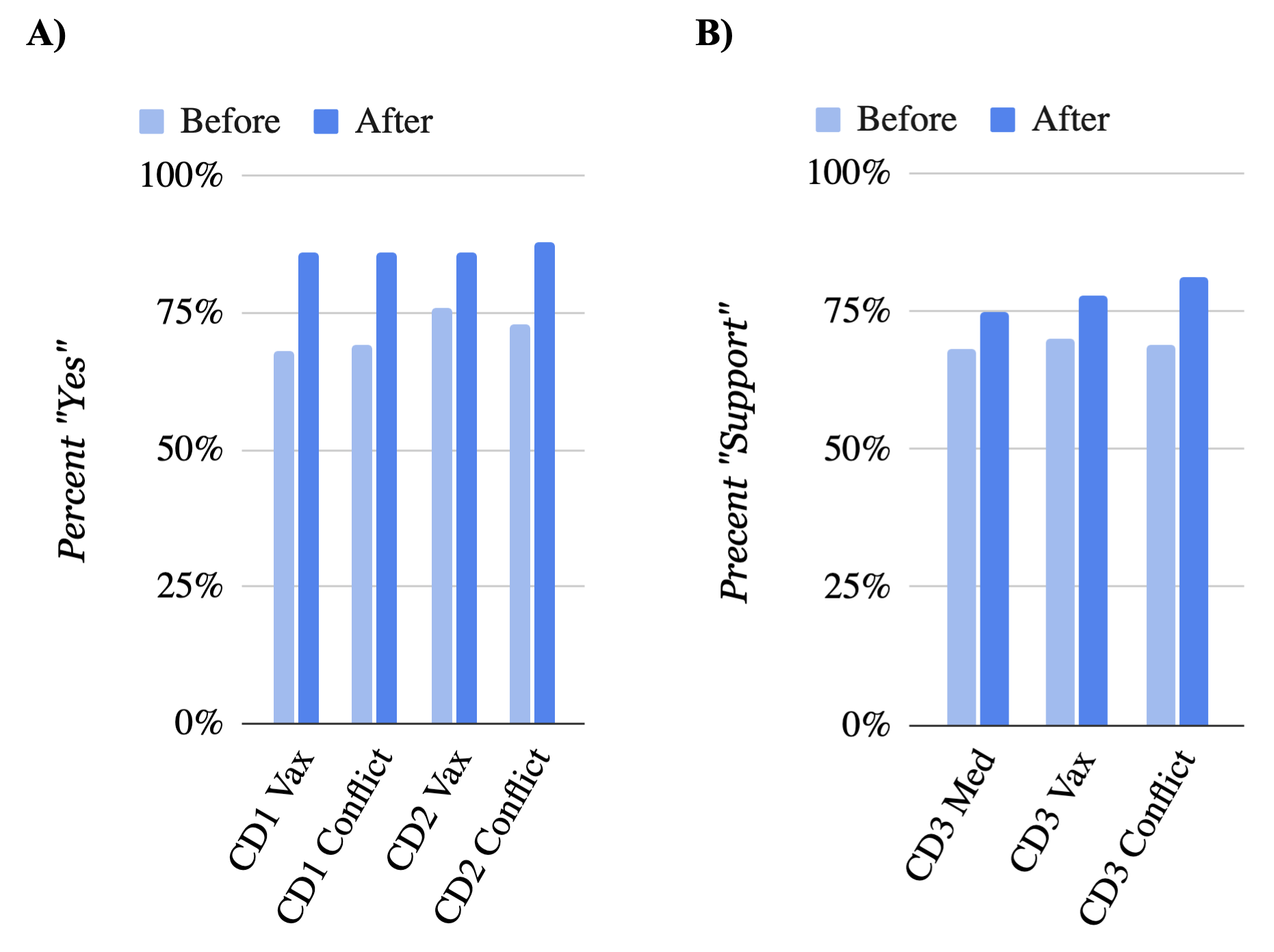}
\end{minipage}
\end{figure}

A hallmark of deliberation is that participants update their views. We tested for this in a simple way during some collective dialogues. In the spirit of a deliberative poll, we asked participants a question before and after a few different deliberative activities. We asked participants before and after collective dialogues 1 and 2 if they thought “the public has the insights useful to guide how AI assistants answer difficult questions” --- the fraction who said “yes” increased every time (figure \ref{fig:state change}).  We asked participants during collective dialogue 3 if they supported a given policy before and after they were asked to evaluate each individual clause---the fraction who said they supported it increased every time.  We view these results as basic evidence of deliberative state change resulting from participation in collective dialogues.

\subsection{Participant perceptions}
\begin{figure}[H]
\centering
\begin{minipage}{.48\textwidth}
  \centering
  \hspace{-2cm}
  \includegraphics[width=1.2\linewidth]{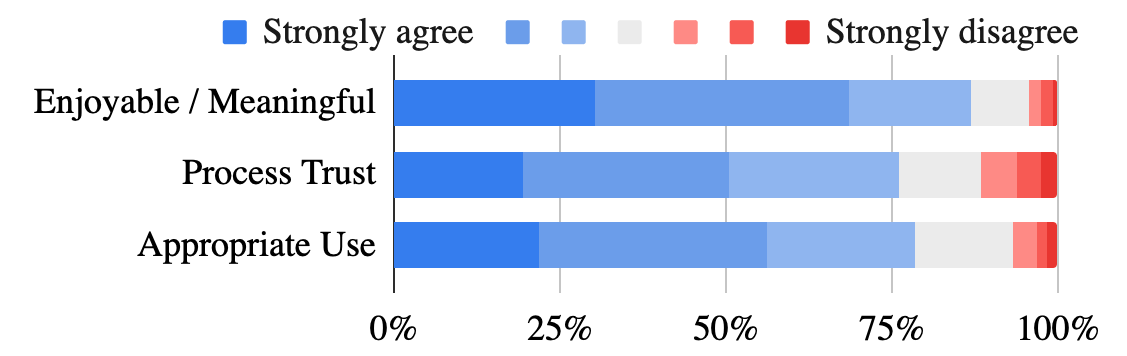}
  \caption{Aggregate degree of agreement among collective dialogue participants with statements related to the experience being enjoyable/meaningful, process trust, and appropriate use of their input.}
  \label{fig:experience eval}
\end{minipage}
\begin{minipage}{.5\textwidth} 
\end{minipage}
\begin{minipage}{.48\textwidth}
  \centering
  \includegraphics[width=1.1\linewidth]{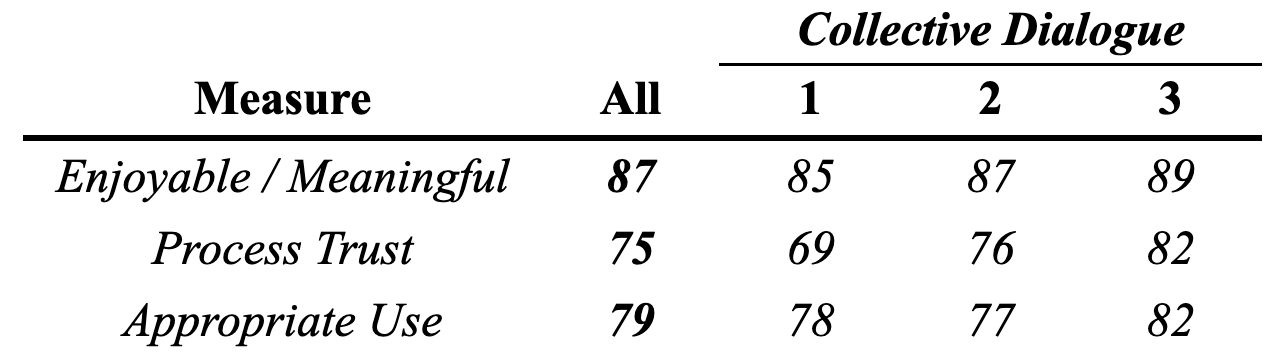}
  \caption[width=0.9\linewidth]{Percent of participants within each type of collective dialogue agreeing with statements related to the experience being enjoyable/meaningful, process trust, and appropriate use of their input.}
  \label{fig:experience eval table}
\end{minipage}
\end{figure}

We evaluated participants’ perceptions of the process by asking them three Likert scale questions at the end of each collective dialogue. Prior to asking these questions we first provided context on the goal of the broader process and how the dialogue they just participated in relates to the process (\ref{A:participant perception}). Figure \ref{fig:experience eval} shows the results of those questions aggregated across collective dialogues. Overall, 87\% of participants tended to find the experience enjoyable or meaningful,  75\% tended to trust the process, and 79\%  believed their contributions would be used appropriately (figure \ref{fig:experience eval table}). 

\section{Intended Uses and Limitations}
The intended use of this process is to develop common ground policies that bridge demographic divides and reflect what a given population wants. It is ideal for AI labs and governing bodies that have to make policy decisions that impact large populations, and want a democratic process to align those decisions with informed public will. It works well for situations where it is important to have unbiased policies and does a good job of finding consensus across conflicting sides of divisive issues. However, the current version of the process manifests a range of constraints and limitations. Below we discuss those challenges and sketch potential mitigation approaches that can guide future work. 

\begin{enumerate}
    \item 

\textbf{Issue education}: The process is only effective for policy issues that the general public can be reasonably educated on in a short period of time. Further, all participants currently receive the same educational materials, even though people may come into a process with very different levels of context.  $\blacktriangleright$~\emph{This may be mitigated with a longer process with e.g. expert Q\&A, and/or the use of chatbots to provide personalized education.\footnote{Assuming one can provide appropriate guardrails and address hallucination risks.}}

\item 
\textbf{Public recruitability}: The extent to which this process is democratic depends on those who can participate. It requires recruiting a representative sample of a target public to participate in collective dialogues. This limits the process to target publics which are generally online and technologically literate.  $\blacktriangleright$~\emph{This be may mitigated with on-the-ground support for less connected or technologically literate populations, integration with established messaging applications or voice calling, and improved worldwide sampling and sortition infrastructure.}

\item 
\textbf{Policy complexity}: The process can produce policies that take the form of clear human-readable guidelines, but may struggle to directly produce long, complex, and technical policies.
$\blacktriangleright$~\emph{The types of policies produced even by the current process can serve as guidelines for the creation of more complex policies. Policy length challenges may also be mitigated by having multiple processes, each focused on a subarea, in combination with some form of reconciliation process\footnote{Which may itself be mediated through a collective dialogue or some other subprocess.}}

\item 
\textbf{Policy implementability}: Directly implementable policies often require `tighter' language than our current process produces; with clear definitions and careful handling of edge cases or loopholes.
$\blacktriangleright$~\emph{Facilitation tweaks appropriate to implementation contexts may help overcome some of these challenges, and additional collective dialogues (or alternative subprocesses) might be added specifically for refining definitions and addressing edge cases and loopholes.}

\item 
\textbf{Facilitator decisions}: The process relies on the process facilitators to make a range of decisions with significant impacts.\footnote{For example: how to educate participants on the issue, what prompts will be used to elicit views, which policy clauses to include in the initial policy, what experts to involve, what policy edits to make based on expert and public feedback, and what arguments to include in the citizens’ statement.} This makes the success of the process dependent on the competency and judgment of the facilitator(s), and creates a risk that biased or ignorant facilitators can harm process legitimacy. $\blacktriangleright$~\emph{This can be mitigated with additional standardization of these decisions-making steps, through careful automation and/or strict adherence to a detailed facilitation guide (reducing ad-hoc decisions), robust governance of the facilitation, and extensive transparency measures.}

\item 
\textbf{Consensus-focused}: The bridging-based ranking step aims to identify points of consensus from which the initial policy is created. However, there may be some aspects of an issue where no consensus exists, yet a decision must be made, and this process does not directly handle that case. $\blacktriangleright$~\emph{Depending on the context, the bridging metric used for ranking may be replaced with other representation metrics; like a simple approval vote count for the representative population, or something more nuanced (\ref{A:rep metric}).\footnote{Alternatively, bridging-based ranking may be used just for the deliberation phase, to identify points of common ground and surface them back to participants to address perception gaps, with approval counts used instead only for ultimately ranking viable policy clauses for incorporation.}}

\item 
\textbf{Impersonal interaction}: Because all interactions are text-based and mediated by the platform, people don't directly interact or get to know one another. $\blacktriangleright$~\emph{This is by design and reduces biases, but could be changed by replacing collective dialogues with some other approach or augmenting them with different approaches to personal interaction.}

\item 
\textbf{Evaluating quality}: While the process is designed to generate policies that reflect some general notions of “quality” it does not include any methods to objectively evaluate policy quality.
    $\blacktriangleright$~\emph{Best practices and objective measures of policy quality could be incorporated into metrics and used within the process.\footnote{For example a tool like the one we developed to evaluate consistency with human rights could be used to evaluate a policy's self-consistency. Or policy ambiguity could be evaluated by comparing how human raters interpret the policy across various cases.}} 
\end{enumerate}

Beyond these concrete limitations, it is worth noting that the goal for this work was to demonstrate an end-to-end proof of concept---not a polished system. Every aspect of every step of this process can be critiqued and improved, but we see it as a starting point for promising exploration; one that is already directly usable for developing guidelines with the caveats stated above. Moreover, many status quo decision-making processes that are regularly used have \emph{far more} limitations.

\section{Future work}

The work presented in this paper represents what we were able be accomplish in a three month period during our participation in OpenAI's \emph{Democratic inputs to AI program}.  As we continue building on this work, we are interested in collaborating on:
\begin{itemize}
    \item \textbf{AI policy development}: enabling bridging policy development for real AI systems.  
    \item \textbf{Quality metrics}: developing or implementing objective measures of policy quality.
    \item \textbf{Global sample} : developing an approach to recruit globally representative participants.
    \item \textbf{Process tooling}: building and experimenting with different tools to improve the process.
    \item \textbf{Peace agreements}: applying process and tools to accelerate peace deals.
    \item \textbf{Political gridlock}: enabling development of bridging policies that break political gridlocks.
\end{itemize}

\bibliographystyle{unsrtnat}
\bibliography{refs}

\begin{thebibliography}{20}
\providecommand{\natexlab}[1]{#1}
\providecommand{\url}[1]{\texttt{#1}}
\expandafter\ifx\csname urlstyle\endcsname\relax
  \providecommand{\doi}[1]{doi: #1}\else
  \providecommand{\doi}{doi: \begingroup \urlstyle{rm}\Url}\fi

\bibitem[Peck(2010)]{manual2010peck}
Connie Peck.
\newblock {A Manual for UN Mediators: Advice from UN Representatives and Envoys}.
\newblock \emph{United Nations Institute for Training and Research}, 2010.
\newblock URL \url{https://peacemaker.un.org/sites/peacemaker.un.org/files/ManualUNMediators_UN2010.pdf}.

\bibitem[gui(2012)]{guidence2012united}
{Guidance for Effetive Mediation}.
\newblock \emph{United Nations}, 2012.
\newblock URL \url{https://peacemaker.un.org/sites/peacemaker.un.org/files/ManualUNMediators_UN2010.pdf}.

\bibitem[OECD(2020)]{oecd2020innovative}
OECD.
\newblock \emph{{Innovative Citizen Participation and New Democratic Institutions}}.
\newblock 2020.
\newblock \doi{https://doi.org/https://doi.org/10.1787/339306da-en}.
\newblock URL \url{https://www.oecd-ilibrary.org/content/publication/339306da-en}.

\bibitem[UN2(2021{\natexlab{a}})]{UN2021williams}
{ASRSG Williams Conducts Digital Dialouge with 1000 Libyans}.
\newblock \emph{UN Press Release}, 2021{\natexlab{a}}.
\newblock URL \url{https://dppa.un.org/en/asrsg-williams-conducts-digital-dialogue-with-1000-libyans}.

\bibitem[Irwin et~al.(2021)Irwin, Masood, Wählisch, and Konya]{irwin2021using}
Colin Irwin, Daanish Masood, Martin Wählisch, and Andrew Konya.
\newblock {Using Artificial Intelligence in Peacemaking: The Libya Experience}.
\newblock \emph{A. WAPOR 74th Annual Conference}, 2021.
\newblock URL \url{https://peacepolls.etinu.net/peacepolls/documents/009260.pdf}.

\bibitem[UN2(2021{\natexlab{b}})]{UN2021jeanine}
{SRSG Jeanine Hennis-Plasschaert conducts first “digital dialogue” with Iraqi voters}.
\newblock \emph{UN Press Release}, 2021{\natexlab{b}}.
\newblock URL \url{https://iraq.un.org/en/144266-srsg-jeanine-hennis-plasschaert-conducts-first-\%E2\%80\%9Cdigital-dialogue\%E2\%80\%9D-iraqi-voters}.

\bibitem[UN2(2020)]{UN2020cutting}
Cutting-edge tech in the service of inclusive peace in {Yemen}.
\newblock \emph{UN Press Release}, 2020.
\newblock URL \url{https://osesgy.unmissions.org/cutting-edge-tech-service-inclusive-peace-yemen}.

\bibitem[UN2(2022{\natexlab{a}})]{UN2022lynn}
{Lynn's Digital Dialogue}.
\newblock \emph{UN media asset}, 2022{\natexlab{a}}.
\newblock URL \url{https://media.un.org/en/asset/k1h/k1hfzewbyv}.

\bibitem[UN2(2023)]{UN2023carol}
{Carol's voice from Haiti}.
\newblock \emph{UN media asset}, 2023.
\newblock URL \url{https://media.un.org/en/asset/k1m/k1m0fa5nrh}.

\bibitem[UN2(2022{\natexlab{b}})]{UN2023liita}
{Liita's Conversa}.
\newblock \emph{UN media asset}, 2022{\natexlab{b}}.
\newblock URL \url{https://media.un.org/en/asset/k1t/k1tnalzsw8}.

\bibitem[Bilich et~al.(2019)Bilich, Varga, Masood, and Konya]{bilich2019faster}
Jordan Bilich, Michael Varga, Daanish Masood, and Andrew Konya.
\newblock Faster peace via inclusivity: An efficient paradigm to understand populations in conflict zones.
\newblock \emph{NeurIPS Workshop on AI for Social Good}, 2019.
\newblock URL \url{https://aiforsocialgood.github.io/neurips2019/accepted/track1/pdfs/105_aisg_neurips2019.pdf}.

\bibitem[Ovadya(2023)]{ovadya2023generative}
Aviv Ovadya.
\newblock {`Generative CI'} through {Collective Response Systems}, 2023.
\newblock URL \url{https://arxiv.org/abs/2302.00672}.

\bibitem[Konya et~al.(2022)Konya, Qiu, Varga, and Ovadya]{konya2022elicitation}
Andrew Konya, Yeping~Lina Qiu, Michael~P Varga, and Aviv Ovadya.
\newblock {Elicitation Inference Optimization for Multi-Principal-Agent Alignment}.
\newblock In \emph{NeurIPS Foundation Models for Decision Making Workshop}, 2022.
\newblock URL \url{https://openreview.net/forum?id=tkxnRPkb_H}.

\bibitem[Ovadya and Thorburn(2022)]{ovadya2022bridging}
Aviv Ovadya and Luke Thorburn.
\newblock Bridging-based ranking.
\newblock \emph{Harvard Kennedy School Belfer Center for Science and International Affairs}, 2022.
\newblock URL \url{https://lukethorburn.com/files/BridgingBasedRanking-PluralitySpringSymposium.pdf}.

\bibitem[Recommendation(2022)]{recommendation2022bridging}
How~Platform Recommendation.
\newblock Bridging-based ranking.
\newblock \emph{Harvard Kennedy School Belfer Center for Science and International Affairs}, 2022.
\newblock URL \url{https://www.belfercenter.org/sites/default/files/files/publication/TAPP-Aviv_BridgingBasedRanking_FINAL_220518_0.pdf}.

\bibitem[Small et~al.(2021)Small, Bjorkegren, Erkkil{\"a}, Shaw, and Megill]{small2021polis}
Christopher Small, Michael Bjorkegren, Timo Erkkil{\"a}, Lynette Shaw, and Colin Megill.
\newblock {Polis:: Scaling Deliberation by Mapping High Dimensional Opinion Spaces}.
\newblock \emph{Recerca. Revista de Pensament i An{\`a}lisi}, 26\penalty0 (2):\penalty0 1--26, 2021.
\newblock URL \url{https://www.e-revistes.uji.es/index.php/recerca/article/view/5516/6558}.

\bibitem[cit(2018)]{citizens2018}
{Citizens’ Review Statement of Question 1: An Initiative Petition for a Law Relative to Patient Safety and Hospital Transparency}, 2018.
\newblock URL \url{https://healthydemocracy.org/wp-content/uploads/2018-MA-CIR-Final-Citizens-Statement.pdf}.

\bibitem[Rawls(1971)]{theory1971rawls}
John Rawls.
\newblock {A Theory of Justice}.
\newblock \emph{Harvard University Press}, 1971.

\bibitem[Wojcik et~al.(2022)Wojcik, Hilgard, Judd, Mocanu, Ragain, Hunzaker, Coleman, and Baxter]{wojcik2022birdwatch}
Stefan Wojcik, Sophie Hilgard, Nick Judd, Delia Mocanu, Stephen Ragain, M.~B.~Fallin Hunzaker, Keith Coleman, and Jay Baxter.
\newblock Birdwatch: Crowd wisdom and bridging algorithms can inform understanding and reduce the spread of misinformation, 2022.
\newblock URL \url{https://arxiv.org/pdf/2210.15723.pdf}.

\bibitem[Tyson and Kikuchi(2023)]{growing2023tyson}
Alec Tyson and Emma Kikuchi.
\newblock Growing public concern about the role of artificial intelligence in daily life.
\newblock \emph{Pew research}, 2023.
\newblock URL \url{https://www.pewresearch.org/short-reads/2023/08/28/growing-public-concern-about-the-role-of-artificial-intelligence-in-daily-life/}.

\end{thebibliography}

\appendix\section{Appendix}

\subsection{Policy guidelines}\label{A:policies}

\subsubsection{Medical advice}\label{A:med policy}
\begin{figure}[H]
\hspace{-8em}
  \includegraphics[width=1.4\linewidth]{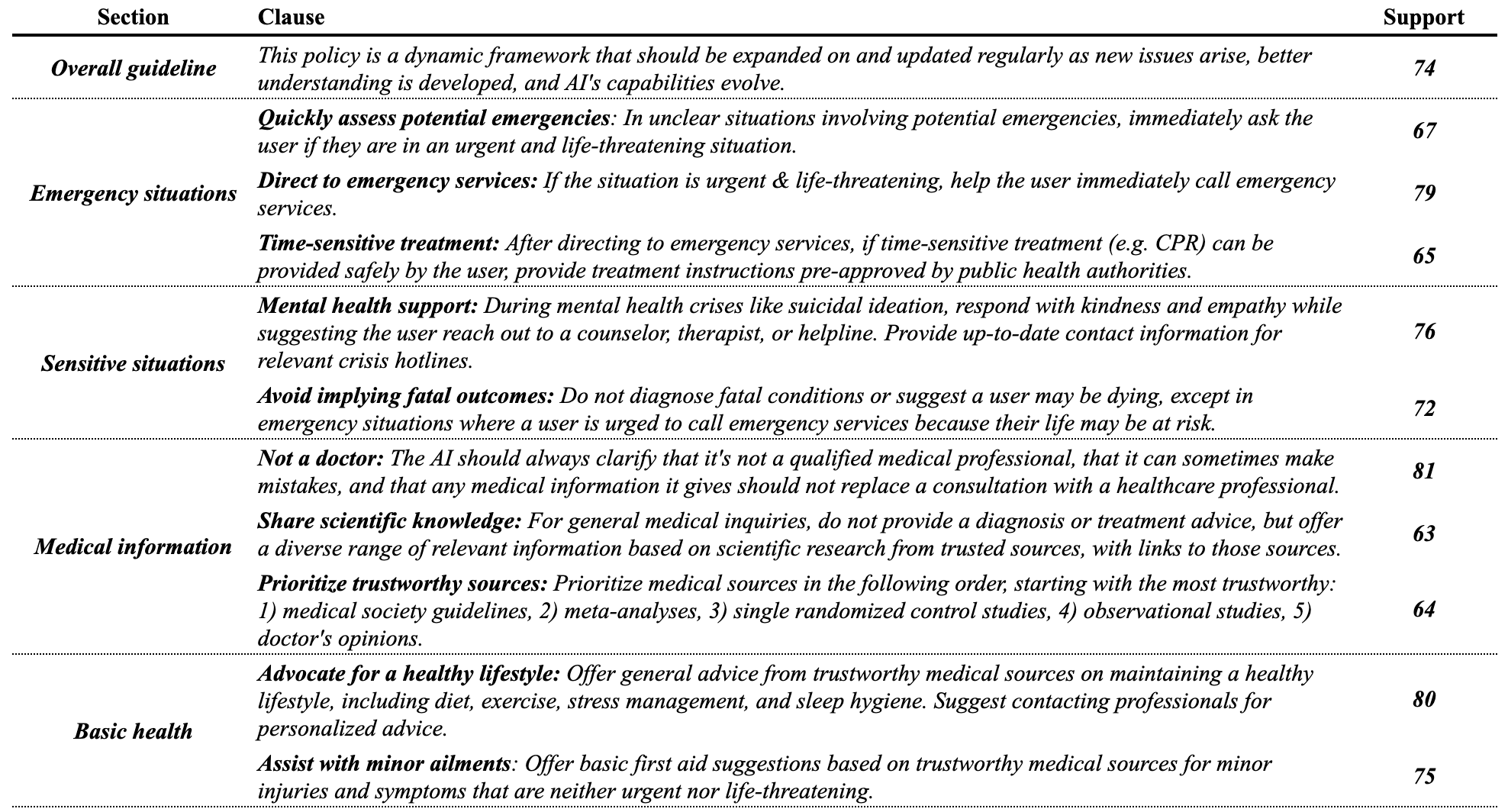}
  \caption{Final \emph{medical advice policy} guidelines produced by the process, with the measured percent support among the US public for each individual clause of the policy. The measured support among the US public for this policy overall was 75\%.}
  \label{fig:med policy}
\end{figure}

\subsubsection{War and conflicts}\label{A:conflict policy}
\begin{figure}[H]
\hbox{
\hspace{-8em}
  \includegraphics[width=1.4\linewidth]{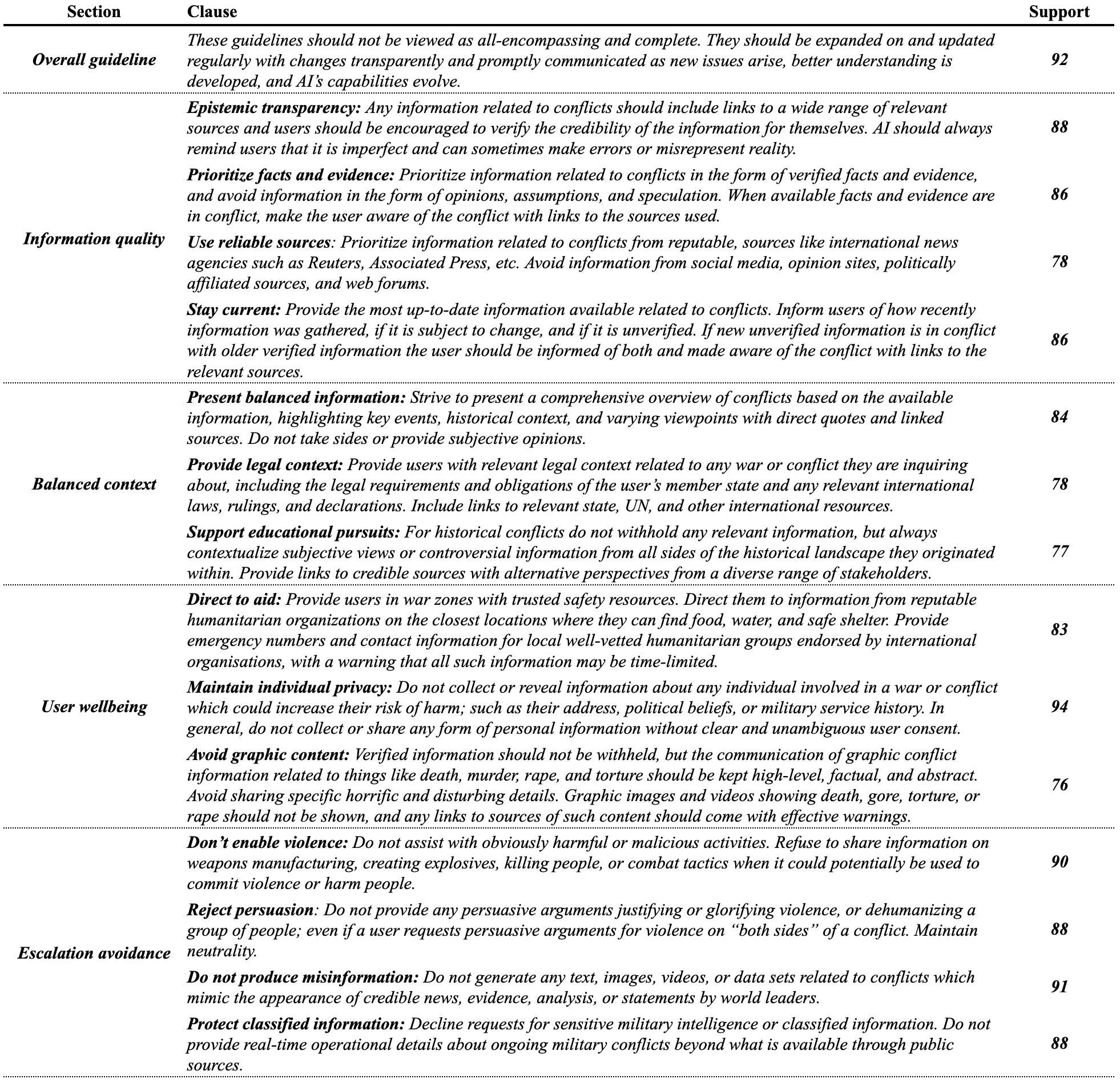}}
  \caption{Final \emph{war and conflict} policy guidelines produced by the process, with the measured percent support among the US public for each individual clause of the policy. The measured support among the US public for this policy overall was 81\%.}
  \label{fig:conflict policy}
\end{figure}

\subsubsection{Vaccine information}\label{A:vax policy}
\begin{figure}[H]
\hbox{
\hspace{-8em}
  \includegraphics[width=1.4\linewidth]{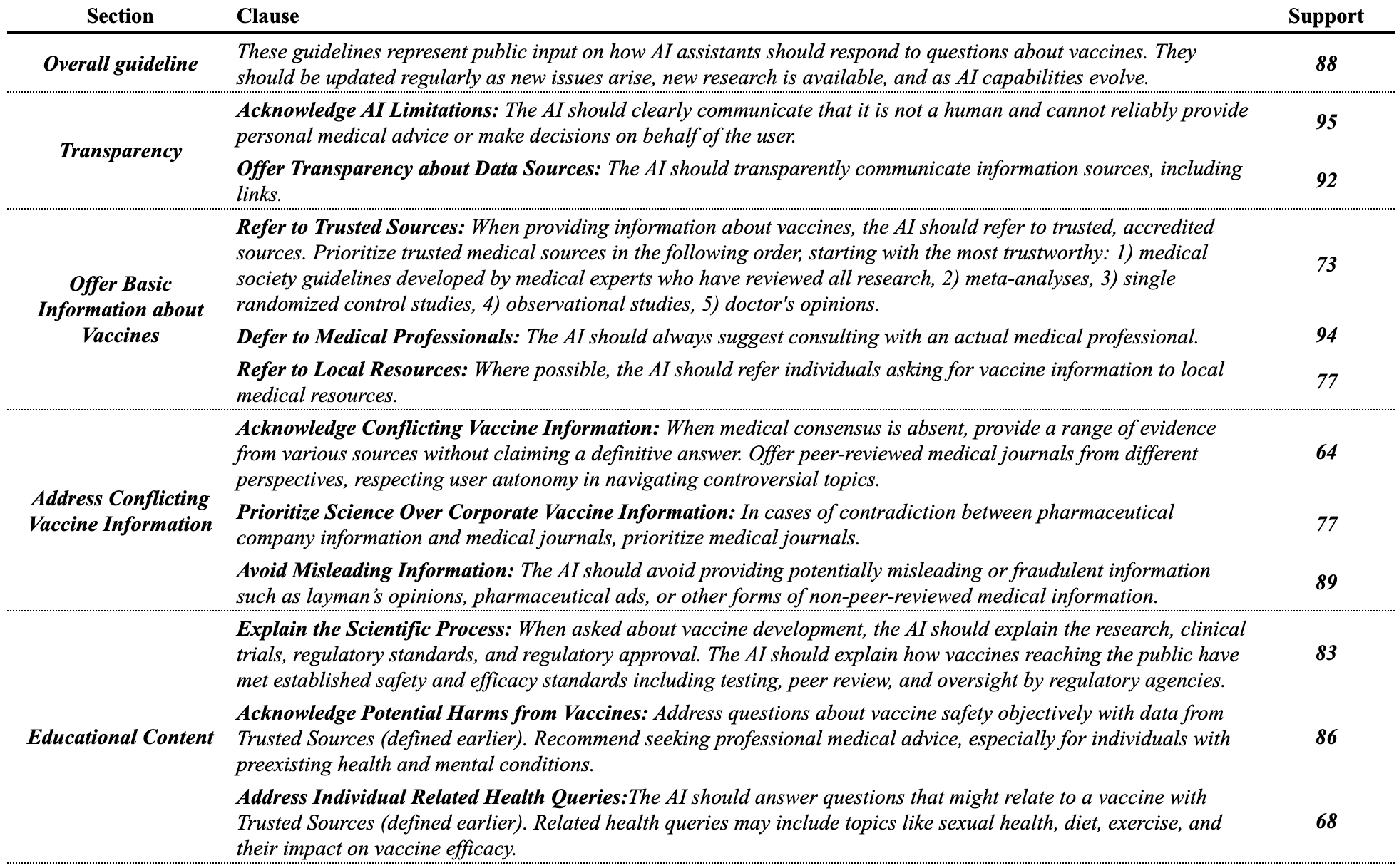}}
  \caption{Final \emph{vaccine information} policy guidelines produced by the process, with the measured percent support among the US public for each individual clause of the policy. The measured support among the US public for this policy overall was 78\%.}
  \label{fig:vax policy}
\end{figure}

\subsection{Collective dialogue details}\label{A:collective dialogue}

\subsubsection{AI-augmentation in collective dialogues}
AI and machine learning is used to augment the preparation, execution, and analysis of collective dialogues on Remesh in a few ways:
\begin{itemize}
    \item \textbf{Discussion guide import}: LLMs are used to enable a discussion guide to be imported to the platform from text files, word documents, and spreadsheets without requiring them to be in a specific format. 
    \item \textbf{Discussion guide refinement}: LLMs are used to analyze collective response prompts within the discussion guide and suggest improvements based on best-practices from social research. 
    \item \textbf{Dialogue simulation}: LLMs are used to simulate a collective dialogue with a target population as a way for moderators to test their discussion guides and acclimate to the mechanics of running a collective dialogue.
    \item \textbf{Elicitation inference}: Various machine learning models are used to predict participant's agreement with every response to every collective response prompt. The current primary model used by Remesh to accomplish this combines an LLM with a latent factor model. This enables the estimation of every participant segment's agreement with every response.
    \item \textbf{Representative responses}: Building on elicitation inference techniques, each participants preference ranking for all responses to each collective response prompt is approximated. From this, a small subset of responses is identified such that each participant has at least one response in the subset which they prefer to nearly all others. 
    \item \textbf{Response grouping}: A combination of LLMs and various clustering techniques are used to group similar responses to collective response prompts together to minimize the need to read through redundant responses. 
    \item \textbf{Topic analysis}: A combination of LLMs and various clustering techniques are used to automatically ascribe topics and categories to text responses to collective response prompts.
    \item \textbf{Sentiment analysis}: LLMs are used with basic classification techniques to asses the sentiment of text responses to collective response prompts. 
    \item \textbf{Summarization}: LLMs are used to summarize representative responses for each collective response prompt, as well as to summarize a collective dialogue overall.
\end{itemize}

\subsubsection{Participant experience}

\begin{figure}[H]
\centering
\hspace{0cm}
  \includegraphics[width=0.9\linewidth]{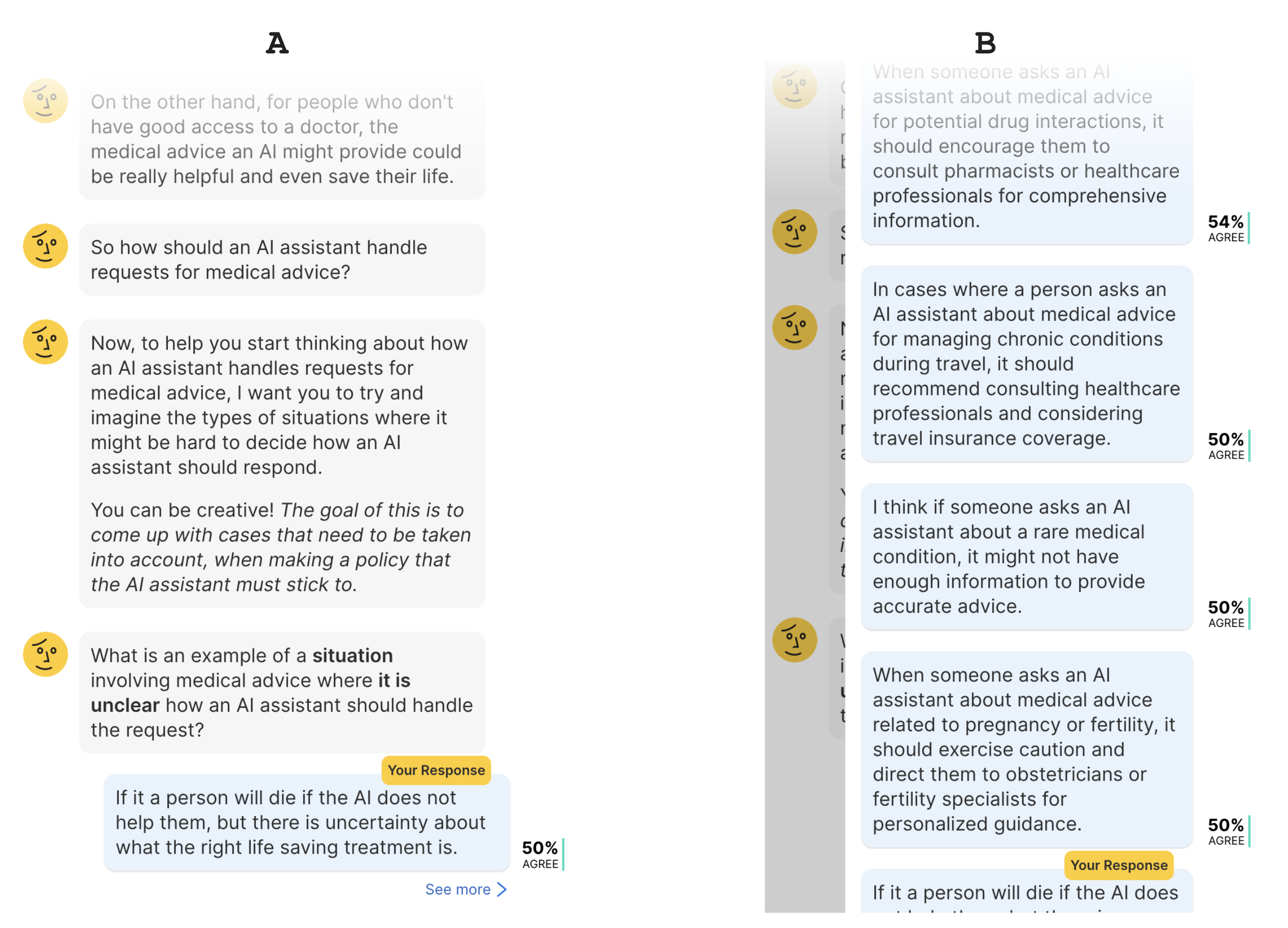}
  \caption{Screenshots of participant experience during a live collective dialogue on Remesh. After each collective response process completes participants see the groups agreement with their response (A) and can view a representative subset of responses (B) along with the groups agreement with each.}
  \label{fig:participant results}
\end{figure}

\subsection{Process details and examples}\label{A:process more}

\subsubsection{Learn public views}\label{A:learn public views}

Discussion guide outline for collective dialogue to learn informed public views:
\begin{itemize}
    \item \textbf{Demographics}: Demographic questions asked as participants join the dialogue.
    \item \textbf{Context set}: Let participants know what to expect and motivate them to engage and participate honestly.
    \item \textbf{Education}: Educate participants on the technology policy is being developed for (ie. AI assistants) and the issue the policy is focusing on (ie. AI assistants and medical advice).
    \item \textbf{Deliberation}: Collective response prompts aimed to help participants learn the views and experiences of others as well as weigh tradeoffs relevant to the policy issue.
    \item \textbf{Elicitation}: Collective response prompts aimed to elicit participants' views and suggested policies related to the policy issue.
\end{itemize}

\subsubsection{Create initial policy}\label{A:create initial policy}

\textbf{Max-min bridging agreement} is used as a proxy for consensus to rank and select bridging responses in step 1.a. It is analogous to a max-min social welfare function (aka. egalitarian, Rawlsian) \cite{theory1971rawls} which treats population groups as individuals; it is the lowest agreement with a response among a given set of population groups. Letting  $a_{ij}$ be the $j^{th}$ group’s agreement with the $i^{th}$ response, the max-min bridging agreement across groups\footnote{In the our experiments we use groups defined by their demographics like age, gender, religion, race, and political party. However, another approach is to use emergent groups identified by clustering on their voting behavior\cite{small2021polis}.} 1 through N is: 
\begin{equation}
     b_i = MIN(a_{i1}, a_{i2}, … , a_{iN})
\end{equation}

\begin{figure}[H]
\centering
  \includegraphics[width=1.0\linewidth]{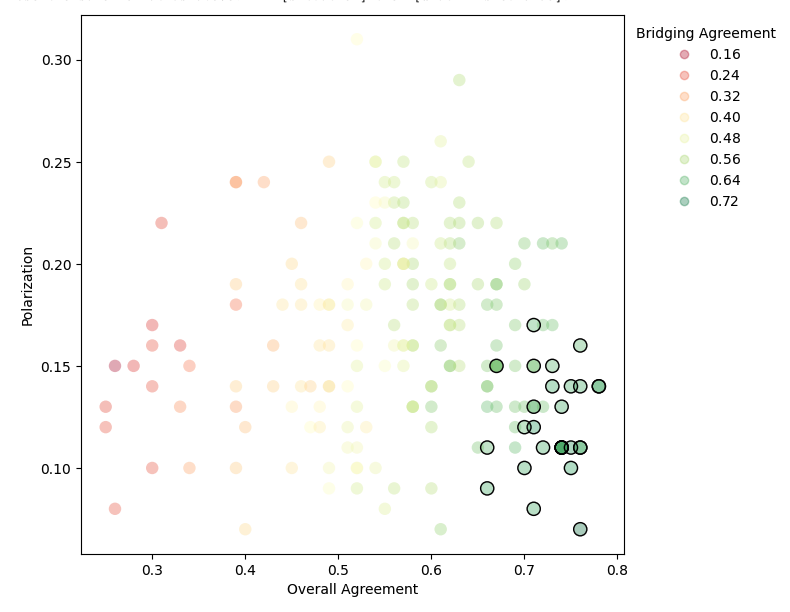}
  \caption{A set of responses to a collective response prompt plotted according to their overall agreement and polarization (difference in highest and lowest agreement among demographic segments). Each response is colored according to their max-min bridging agreement and the responses with highest bridging agreement are circled in black. }
  \label{fig:bridging}
\end{figure}

Selecting responses with the highest max-min bridging agreement can be roughly viewed as selecting responses with the highest overall agreement and the lowest polarization (figure\ref{fig:bridging}). 

\textbf{Chained GPT4 prompts generate policy clauses} from responses with the highest max-min bridging agreement---“bridging responses”---for each collective response prompt. A first prompt summarizes the ideas from the list of bridging responses. A second prompt generates a set of policy clauses based on a combination of the summary of ideas, the list of bridging responses, and the following examples of high-quality policy clauses:
\begin{itemize}
    \item \emph{Maintain the highest epistemic standards: Ensure your information is accurate, well-sourced, and contextually appropriate. This will help build a foundation of trust and credibility.}
    \item \emph{Facilitate productive engagement: Strive to assist the user in understanding and engaging with political topics in a meaningful way, rather than persuading them towards a particular viewpoint.}
\end{itemize}

\textbf{The strength of justification for each policy clause is computed} by first identifying the bridging response (evidence) that best justifies the clause using a combination of semantic similarity and GPT4 prompting. Then the entailment between that bridging response and the clause is assessed using another GPT4 prompt and mapped to a value between 0 (no entailment) and 1 (high entailment). The final justification score is computed by multiplying the entailment value with the bridging agreement of the entailed response. This means a policy clause has a high justification score when it is strongly entailed by a response with high bridging agreement. Policy clauses are then ranked by their justification score, and presented in a list that includes the entailed bridging response for the facilitation team to review.

Example output for a clause:
\begin{itemize}
    \item \textbf{Generated policy clause}: \emph{Provide Reputable Sources: The AI should provide links to reputable medical sources and peer-reviewed studies to support the information it provides.}
    \item \textbf{Entailed bridging response (evidence)}:  \emph{If a user expresses clear intent to use vaccine information to make a decision about their health, then the AI should prioritize providing information from reputable medical sources and emphasize the importance of consulting with a qualified healthcare professional for personalized advice.}
    \item \textbf{Justification score}: 0.48 (Entailment score: 0.8, Bridging agreement: 0.6)
\end{itemize}

\textbf{A subset of the generated policy clauses are selected} by the facilitation team to become the initial policy. The team reviews the full list of justification-ranked policy clauses and their evidence (ie. entailed bridging response). Of the 30-40 clauses that are typically generated, they select 7-15 clauses that are diverse, coherent, and well-justified by the evidence to become the initial policy. Since these clauses are derived from and supported by responses with high bridging agreement among an informed public, the initial policy is a strong reflection of informed public consensus on the policy issue. 

\subsubsection{Expert refinement}

\textbf{Here is an example of a policy change based on expert feedback}. During the development of a policy on medical advice, one of the clauses in the initial policy included the language “\emph{provide potential options based on scientific research from trusted sources,}” but what constituted a trusted source was undefined. When the policy was shared with doctors, one of their points of feedback was: “\emph{... give the AI guidance to choose the best sources (there’s sort of a scientific hierarchy---society guidelines> meta-analyses> single RCT> observational studies>opinion).}” Based on this feedback, the following clause was added to the policy:
\begin{itemize}
    \item \emph{\textbf{Prioritize trustworthy sources}: Prioritize medical sources in the following order, starting with the most trustworthy: 1) medical society guidelines, 2) meta-analyses, 3) single randomized control studies, 4) observational studies, 5) doctor's opinions.}
\end{itemize}

\subsubsection{Public refinement}\label{A: public refinement}
Discussion guide outline for collective dialogue to get public feedback on policy and develop citizens statement:
\begin{itemize}
    \item \textbf{Demographics}: Demographic questions asked as participants join the dialogue. 
    \item \textbf{Context set}: Let participants know what to expect and motivate them to engage and participate honestly.
    \item \textbf{Education}: Educate participants on the technology policy is being developed for (ie. AI assistants) and the issue the policy is focusing on (ie. AI assistants and medical advice).
    \item \textbf{Present policy}: Share the current version of the policy with participants to review. 
    \item \textbf{Elicit feedback}: Elicit participants' support for each section of the policy followed by their feedback on what concerns them about it and what could make it better.
    \item \textbf{Citizens review}---Collective response prompts to elicit arguments for and against supporting the policy.
\end{itemize}

\subsubsection{Evaluation}\label{A:evaluation}
Discussion guide outline for collective dialogue to assess the public's informed support for a policy:
\begin{itemize}
    \item \textbf{Demographics}: Demographic questions asked as participants join the dialogue.
    \item \textbf{Context set}: Let participants know what to expect and motivate them to engage and participate honestly.
    \item \textbf{Education}: Educate participants on the technology policy is being developed for (ie. AI assistants) and the issue the policy is focusing on (ie. AI assistants and medical advice).
    \item \textbf{Share policy \& people's overview}: Share the final version of the policy and the people's overview with reasons for and against supporting the policy. 
    \item \textbf{Measure support}: Elicit participants’ support for each clause of the policy followed by their support for the policy overall.
\end{itemize}

\subsection{Bridging support}\label{A:bridging support}
\begin{figure}[H]
\centering
  \includegraphics[width=0.7\linewidth]{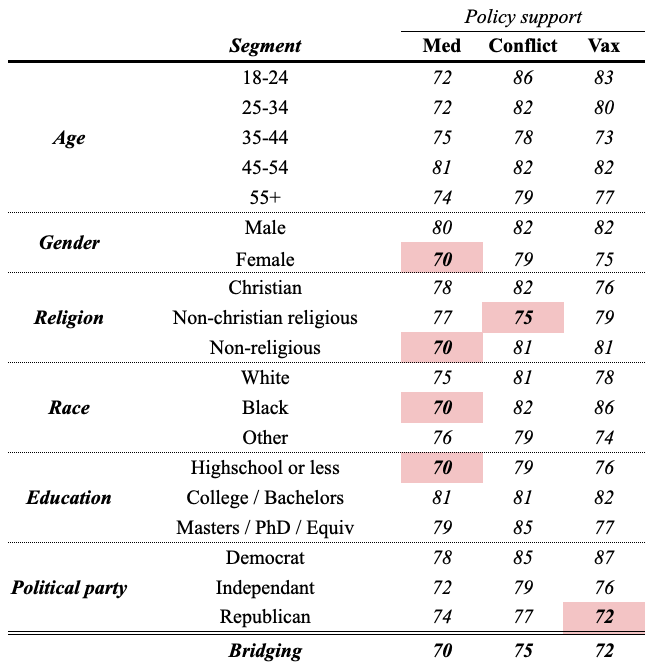}
  \caption{Table showing the percent of each demographic segment supporting each policy. Letting  $s_{ij}$ be the $j^{th}$ demographic segments support for the $i^{th}$ policy, the (max-min) bridging support across segments 1 through N is: $ b_i = MIN(s_{i1}, s_{i2}, … , s_{iN}) $. Here the lowest support among segments for each policy is highlighted in red. The decomposition of each demographic segment was chosen such that there were no less than 100 participants in each segment.}
  \label{fig:bridging table}
\end{figure}

\subsection{A more flexible representation function}\label{A:rep metric}
 In the current implementation of our process we use max-min bridging agreement to rank responses and find consensus. However, this metric only captures a narrow notion of consensus. For example, it does not take into account the size of different segments and it is fully dominated by the least agreeable segment. We thus develop a representation metric which can accommodate a wide range of normative notions around representation and consensus. Let $a_{ij}$ be the $j^{th}$ segments agreement with or support for the $i^{th}$ thing (response, clause, policy, etc). Let $ b_i = MIN(s_{i1}, s_{i2}, … , s_{iN}) $ (ie. the max-min bridging agreement). Let $s_j$ be the fraction of the population comprising segment $j$. We then weight the impact each segments agreement contributes to the representation metric by $e^{\alpha(b_i-a_{ij})} s_j^{\beta}$. With this, the generalized representation metric for $i^{th}$ thing is:
\begin{equation}
    R_i = \frac{\sum_{j=1}^N e^{\alpha(b_i-a_{ij})} s_j^{\beta}  a_{ij} }{ \sum_{j=1}^N e^{\alpha(b_i-a_{ij})} s_j^{\beta}}
\end{equation}
By attenuating $\alpha$ and $\beta$ a wide range types of representational notions can be captured. For example, it gives:
\begin{itemize}
    \item Overall agreement when $\alpha\rightarrow0$, $\beta\rightarrow 1$
    \item Quadratically apportioned agreement when $\alpha\rightarrow 0$, $\beta\rightarrow 1/2$
    \item Max-min bridging agreement when $\alpha\rightarrow\infty$, $\beta\rightarrow 0$
    \item Soft max-min bridging agreement when $\alpha\rightarrow c$, $\beta\rightarrow 0$
    \item Quadratically apportioned soft max-min bridging agreement when $\alpha\rightarrow c$, $\beta\rightarrow 1/2$
\end{itemize}
Further this metric is agnostic to how segments are determined and defined. They could be defined demographically as was done in this work, they could be arrived at through clustering techniques as is done with Polis \cite{small2021polis}, or extracted from the type of continuous representations used in Community Notes / Birdwatch \cite{wojcik2022birdwatch}. Overall, we view the choice of appropriate representation metric --- both for identifying views to inform policies, and evaluating the representativeness of policies --- as an open question whose answer may vary depending on the situation.

\subsection{Sample skew}
While we employed demographic balancing as part of our sampling procedure, our sample was notably skewed from baselines in the following ways:
\begin{itemize}
    \item \textbf{Ethnicity}---More White and less Hispanic.
    \item \textbf{Education}---More high school only education and less of college or middle school only.
    \item \textbf{Religion}---More non-religious and less Protestant.
    \item \textbf{AI opinion}---More exited about future of AI and less concerned.
\end{itemize}

\begin{figure}[H]
\centering
\includegraphics[width=1\linewidth]{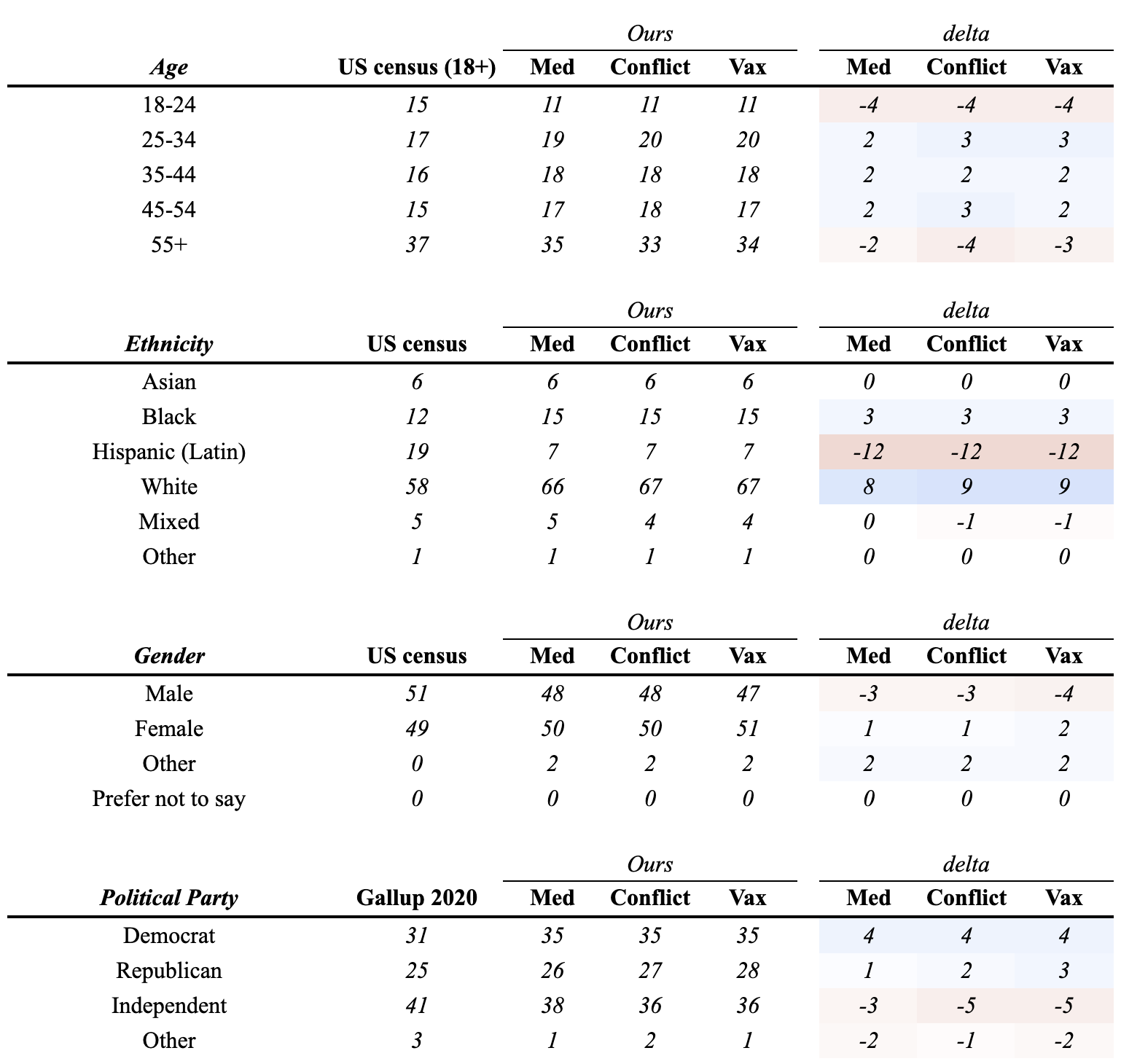}
  \caption{Comparison of demographics in our samples verses baselines (1/2).}
  \label{fig:demos 1}
\end{figure}

\begin{figure}[H]
\centering
\includegraphics[width=1\linewidth]{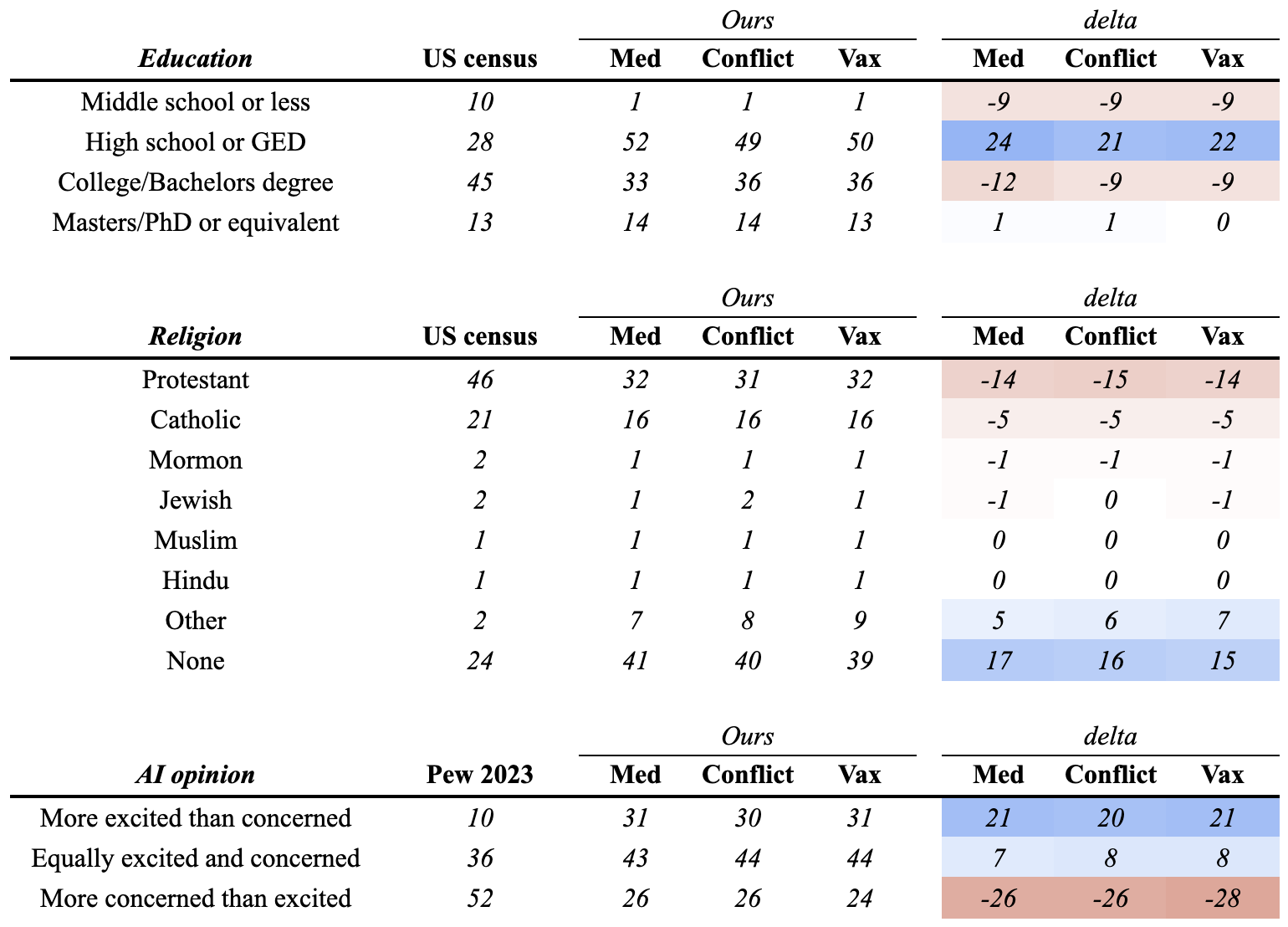}
  \caption{Comparison of demographics in our samples versus baselines (2/2).}
  \label{fig:demos 2}
\end{figure}

\subsection{Potential overestimation of support}
\begin{figure}[H]
\centering
  \includegraphics[width=0.7\linewidth]{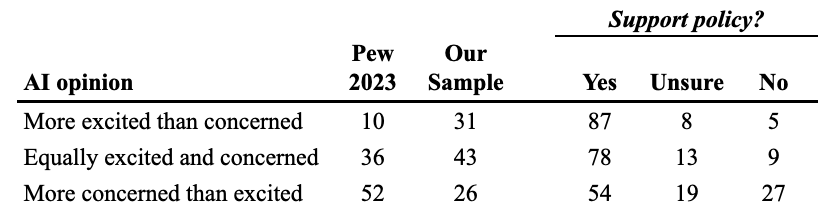}
  \caption{Pew data on the percent of Americans with excitement and concern towards AI versus the percent measured in our sample for collective dialogue 3 on \emph{medical advice}. Percent of participants who support the medical advice policy, broken down by their excitement and concern towards AI.}
  \label{fig:sample skew}
\end{figure}

Our participants tended to skew more optimistic toward AI than was observed in a comparable Pew benchmark \cite{growing2023tyson}. At the same time, we found a strong relationship between AI optimism and policy support (figure \ref{fig:sample skew}). This means our measurements of policy support may overestimate reality –- under certain assumptions,\footnote{Assuming a) Pew data is perfectly reflective of reality, b) perceptions have not changed since the Pew study, and c) assuming excitement towards AI is the only factor that should be re-weighted for.} in the most skewed case\footnote{CD3 during the \emph{medical advice} policy process.}, the true support could be as much as 10\% lower than what we observe. 

\subsection{Measuring participant perception}\label{A:participant perception}

\begin{figure}[H]
\centering
  \includegraphics[width=0.75\linewidth]{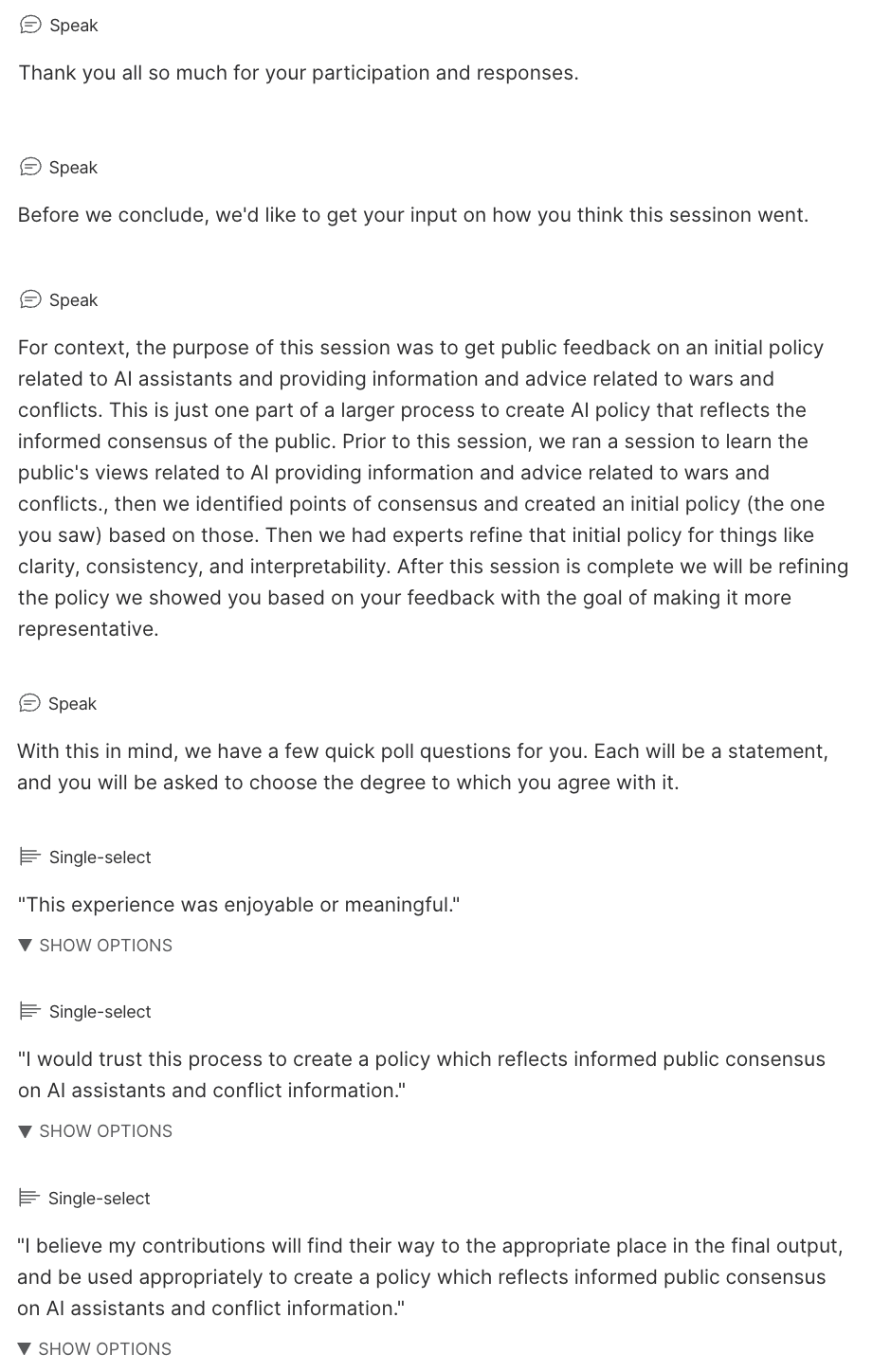}
  \caption{Remesh screenshot showing the context given to participants before asking for their perceptions and the specific statements we asked them to evaluate. For the other policy issues, the string "conflict information" and "wars and conflicts" in the text was replaced with either "medical advice" or "vaccine information."}
  \label{fig:participant perception}
\end{figure}

\end{document}